\documentclass[aps,showpacs,superscriptaddress,nofootinbib,amsmath,amssymb, 10pt]{revtex4}
 \usepackage{graphicx}% Include figure files
\usepackage{dcolumn}% Align table columns on decimal point
\usepackage{bm}% bold math
\usepackage{lipsum}
\usepackage{adjustbox}
\usepackage{multirow}
\usepackage{amsmath}
\usepackage{amssymb}
\usepackage{enumerate}

\graphicspath{./fig}
\begin{document}
% \title{Nuclear medium effects in $F_{1A}  (x,Q^2)$, $F_{2A}  (x,Q^2)$ and $F_{LA}  (x,Q^2)$ structure functions}
 \title{Theoretical modeling of charged current $\nu_\mu(\bar\nu_\mu)-^{40}Ar$ DIS at DUNE energies}
\author{F. Zaidi}
\affiliation{Department of Physics, Aligarh Muslim University, Aligarh - 202002, India}
\author{S. Akhter}
\author{M. Sajjad Athar\footnote{Corresponding author: sajathar@gmail.com}}
\author{S. K. Singh}
\affiliation{Department of Physics, Aligarh Muslim University, Aligarh - 202002, India}
\begin{abstract}
The charged current $\nu_\mu(\bar{\nu}_\mu)$-induced deep inelastic scattering (DIS) from an $^{40}\mathrm{Ar}$ target 
is studied using a microscopic framework that incorporates nuclear medium effects due to Fermi motion, 
binding energy, nucleon correlations, mesonic ($\pi$ and $\rho$) contributions, and nuclear shadowing and antishadowing 
across the relevant Bjorken-$x$ region. The nuclear structure functions $F_{iA}(x,Q^2)$ $(i=1\text{--}3)$ are evaluated using a relativistic nucleon 
spectral function ($S_h$) within the local density approximation employing the free nucleon structure functions, $F_{iN}(x,Q^2)$ $(i=1\text{--}3)$. These 
$F_{iN}(x,Q^2)$ $(i=1\text{--}3)$ are calculated using parton distribution functions (PDFs)
from MMHT 2014 parameterization, including higher-order perturbative QCD corrections up to next-to-next-to-leading order (NNLO), along with 
nonperturbative target mass corrections (TMC). 
% These are subsequently convoluted with the spectral function to obtain the 
% corresponding nuclear structure functions, $F_{iA}(x,Q^2)$, thereby embedding nuclear effects in a self-consistent manner.
The resulting nuclear structure functions $F_{iA}(x,Q^2)$ $(i=1\text{--}3)$ are subsequently used to compute the differential DIS cross sections for 
$^{40}Ar$ nucleus. Numerical results are presented for $\nu_\mu(\bar\nu_\mu)$ beam energies $E=4$ GeV and $E=6$ GeV for the differential scattering
cross sections $\frac{d^2\sigma}{dx dy}$ and $\frac{d\sigma}{dx}$, relevant to ongoing and upcoming liquid-argon neutrino experiments such as 
DUNE and the Fermilab Short-Baseline Neutrino program.
  \end{abstract}
\maketitle
 
 \section{Introduction} 
 
 The study of (anti)neutrino interaction cross sections off free nucleon as well as nucleons bound inside a nuclear target is required to be made
 with better theoretical understanding and with improved statistics in the experimental measurements in order to minimize the present uncertainties in the extraction of the 
 oscillation parameters, viz., mixing angles $\theta_{12}$, $\theta_{23}$, $\theta_{13}$, the Dirac CP-violating phase $\delta_{\mathrm{CP}}$, the
mass-squared differences $\Delta m_{21}^2$ and $|\Delta m_{31}^2|$, etc., to a few percent level, and to resolve the neutrino mass hierarchy problem. Since the (anti)neutrino-nucleon/nucleus cross sections
are small, all the neutrino experiments in general require to employ large volume detectors filled with medium-to-heavy nuclear targets like carbon, oxygen, argon,
iron, lead, etc. For example, the next-generation experiments such as the Deep Underground Neutrino Experiment (DUNE) at Fermilab in USA~\cite{DUNE:2018tke, Cicero:2025mab} and
the Hyper-Kamiokande (HyperK) in Japan~\cite{Hyper-Kamiokande:2025fci} which are designed to measure the CP violating 
phase $\delta_{CP}$ in the leptonic sector, neutrino mass hierarchy, matter-anti matter asymmetry, etc., with enhanced precision. These neutrino experiments
are planned with large nuclear targets such as the multi-kiloton liquid-argon time projection chambers (LArTPCs) in DUNE
($\langle E\rangle=2.5-3$ GeV) and 260,000 metric tons of ultrapure water in HyperK ($\langle E\rangle=0.6$ GeV), and 
focus specifically on understanding the (anti)neutrino-nucleus cross sections in the few GeV energy region which is sensitive to some of the oscillation parameters. Moreover, the Short-Baseline
 Neutrino (SBN) program$-$comprising ICARUS~\cite{ICARUS:2023gpo, Wood:2024jos}, SBND~\cite{SBND:2025lha, Freire:2025hvi}, and 
 MicroBooNE~\cite{MicroBooNE:2025aiw} experiments$-$employ liquid-argon time projection chamber (LArTPC) technology
 to perform high-resolution measurements of neutrino-argon interaction cross sections in the few-GeV energy regime. Hence, the understanding of (anti)neutrino-nucleus 
 interactions in this energy region is important in order to interpret the presently available and the upcoming data from the aforementioned experiments.

In the few-GeV energy region, (anti)neutrino-nucleon scattering cross sections receive contributions
from the quasi-elastic (QE), the inelastic (Inel), and the deep inelastic scattering (DIS) processes and provide an important probe 
of nucleon structure over a wide range of four momentum transfer square $Q^2$ and $W$ the invariant hadronic mass. The 
(anti)neutrino-nucleon cross sections for the inelastic scattering are generally described in terms of the cross sections arising due to the 
resonant($P_{33}(1232)$, $P_{11}(1440)$, $S_{11}(1535)$, etc.) production amplitude, non-resonant production amplitude and interference between them. 
In this resonance region, $\nu_l(\bar\nu_l); ~l=e,\mu$ interactions with the nucleon target produce various final states with mesons
like $\pi N$, $\pi\pi N$, $\eta N$, $\Lambda K$, etc. states. In the case of resonance production, the vector form factors, for the $N\to R$ transition 
are used which are derived using helicity amplitudes, determined from the real and/or virtual photon scattering experiments~\cite{SajjadAthar:2022pjt}.
In case of the $N\to R$ axial-vector transition form factors, the information is very limited except for the $\Delta(1232)$ for which
there are different parameterizations. It has been strongly felt that precise $\nu_l(\bar\nu_l)-N$ scattering 
cross sections in the $\Delta$-resonance region is required. The description of (anti)neutrino-nucleon scattering in this kinematic
region becomes progressively 
uncertain at higher invariant masses, where several overlapping resonances contribute and experimental constraints on the 
transition form factors for the $N \to R$ transitions remain limited. 

In the region of high $Q^2$ and $W$, the dominant contribution to the scattering cross sections comes from the DIS region, where the interaction is described 
in terms of the incoherent scattering of (anti)neutrinos from partonic constituents of the nucleon. In this regime, perturbative QCD 
(pQCD) offers a well established theoretical framework, in which the structure functions are expressed in terms of the 
parton distribution functions (PDFs). However, the current generation of (anti)neutrino experiments which are using (anti)neutrino beam in the region 
of a few GeV energies like 
NO$\nu$A~\cite{Choudhary:2026cjh, Kalitkina:2025hbg}, MINER$\nu$A~\cite{MINERvA:2025hzq, MINERvA:2023ner}, 
ArgoNeuT~\cite{Duffy:2021hie, ArgoNeuT:2020kir}, etc., and the next generation experiments like DUNE and HyperK, etc., most of the events do not lie inside 
the strict kinematic domain of the applicability of the pQCD. Particularly, the kinematic region of moderate 
$W$, i.e., $W\lesssim 2~ GeV$ and $Q^2\lesssim 2~GeV^2$ are important which lie in the transition region between the resonance
production and the deep inelastic scattering. In this domain of $Q^2$ and $W$, the nonperturbative QCD effects like the target mass corrections, higher twist effects, multi-parton correlations, etc., become
important. 
% Due to the lack of a comprehensive theoretical framework for describing the resonance-DIS transition region, it has become 
% common in phenomenological studies to 
% extend the DIS based prescriptions to also include the region of the lower values of $Q^2$ and $W$. This provides a practical 
% and consistent description 
% of the inclusive cross sections across a broad kinematical range. 
% Nevertheless, one has to be careful in extrapolating the DIS description in the 
% kinematic range to low $Q^2$ and $W$,
% and thus the role of kinematical constrain on $W$ becomes important in the DIS cross sections like $W\ge 1.7$ GeV or $W\ge 2$ GeV is applied 
% to exclude the region which is theoretically understood to be described by the contribution from the excitation of higher
% resonances~\cite{SajjadAthar:2022pjt, SajjadAthar:2020nvy}.
Due to the lack of a comprehensive theoretical framework for describing the resonance to DIS transition region, it has become common in 
phenomenological studies to extend DIS-based prescriptions to include regions of lower values of $Q^2$ and $W$. This provides a 
practical and consistent description of inclusive cross sections across a broad kinematical range. Nevertheless, one must exercise 
caution when extrapolating the DIS description to low  $Q^2$ and $W$. In this context, kinematical constraints on $W$ become important; 
for example, cuts such as $ W \ge$1.7 GeV or $W \ge$ 2 GeV are applied to exclude 
regions that are theoretically understood to be dominated by contributions from the excitation of higher resonances~\cite{SajjadAthar:2022pjt, SajjadAthar:2020nvy}.

\begin{figure}
  \includegraphics[height=6 cm, width=10 cm]{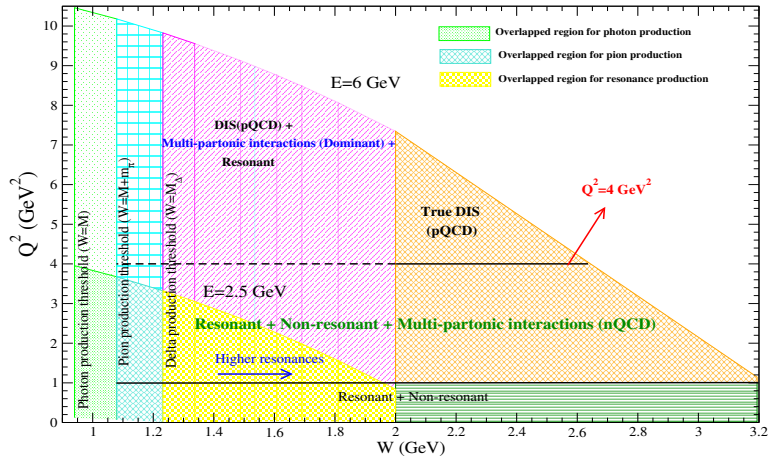}
  \caption{$W$-$Q^2$ plane showing the indicative kinematic boundaries for various possible processes 
at the different neutrino energies viz. $E=2.5$ GeV and $E=6$ GeV~\cite{SajjadAthar:2025fhk}.}
\label{figk}
 \end{figure}

In Fig.~\ref{figk}, we have indicated the
kinematically allowed regions in $W$-$Q^2$ plane for the resonant, non-resonant, and the deep inelastic scattering processes. 
We broadly classify interactions into traditional DIS and shallow inelastic scattering (SIS) region. 
It is important to point out that the direct meson production in the non-resonant region and the meson production 
through the quark-quark interactions (not describable
by the perturbative QCD) at energies lower than the SIS region intermix with the resonant meson production 
and there is no possible way to separate experimentally, individual events that include mesons produced 
by these processes. Therefore, in practice, it is difficult to experimentally have a well-defined SIS region,
clearly separated from the perturbative DIS region or the resonance region. The physics of this complex
region has to be understood better both theoretically and experimentally. The current neutrino experimental community often 
takes $Q^2 \ge 1$ GeV$^2$ and $W\ge 2$ GeV to define the ``DIS'' region~\cite{MINERvA:2025hzq}. These limits on $Q^2$ and $W$ are below the
conventional limits, generally taken for defining a clean DIS region to be described by the perturbative QCD.
This phenomenological separation of events between the DIS and SIS reflect distinct theoretical and experimental challenges. Both kinematic 
regions are highly relevant for DUNE, where over 50$\%$ of events are expected to come from the kinematic region of higher 
resonance excitations and DIS~\cite{DUNE:2018tke, Cicero:2025mab}. 
In oscillation experiments, mis-modeling of the interaction cross sections in the transition region directly impacts neutrino energy reconstruction 
and contributes substantially to the systematic uncertainties. Reducing these uncertainties is essential for achieving the precision goals 
of DUNE, particularly in measurements of $\delta_{\mathrm{CP}}$ and the neutrino mass ordering.

Till date, substantial effort has been devoted 
to improve the neutrino event generators, which are indispensable tools for the experimental analyses and work
as a bridge between the theoretical and experimental sectors. Presently, GENIE MC~\cite{GENIE:2021npt} is the most widely used neutrino event generator
which employs leading-order (LO) parton model cross sections with phenomenological corrections, 
based on the Bodek-Yang model~\cite{Bodek:2010km} for obtaining the DIS cross sections, and the transition region is handled through the empirical prescriptions
in invariant mass $W$ with several limitations as discussed in Ref.~\cite{SajjadAthar:2020nvy}. In the kinematic region of low-to-moderate $Q^2$ and high $x$, 
the separation between DIS and non-DIS processes becomes blurred, 
and partonic descriptions must be supplemented by nonperturbative corrections. These effects are particularly relevant 
for DUNE, where a substantial fraction of events populate the kinematic region of non-DIS events. 
Therefore, a better theoretical understanding of the (anti)neutrino-nucleon/nucleus deep inelastic scattering cross sections, and their modifications
in presence of the higher order perturbative and nonperturbative QCD corrections, the nuclear medium effects  
as well as the kinematic phase-space constraints such as a cut on the hadronic invariant mass (center-of-mass energy) $W$ and the
four momentum transfer square ($Q^2$) is required. In the electromagnetic sector, the importance of nuclear medium effects in the DIS region was clearly established by
the European Muon Collaboration (EMC)~\cite{EuropeanMuon:1983wih} at CERN. It was observed that the ratio of the nucleon structure functions for iron to 
deuterium, $\frac{F_2^N(Fe)}{F_2^N(D)}\;\ne \;1$. This is known in literature as the EMC effect.
Later, in many more experiments at SLAC, CERN and FNAL, similar observations were made using different nuclei~\cite{Arneodo:1992wf, Gomez:1993ri}. 
These nuclear medium effects exhibit a clear dependence on Bjorken-$x$. At large $x (\gtrsim 0.7)$, they are predominantly governed by Fermi motion; in the
intermediate region $0.3 \le x\le 0.7$, they arise from nucleon-nucleon correlations and mesonic contributions-collectively known as the EMC effect; while at
low $x$, they are driven by coherent nuclear phenomena, namely shadowing($x\le 0.1$) and  antishadowing($0.1\le x\le 0.3$). In (anti)neutrino-nucleus 
scattering, however, the interaction involves not only a vector component but also an axial-vector contribution, implying that the corresponding nuclear 
medium effects need not be identical in the two cases. Therefore, an independent understanding of nuclear medium effects in the weak sector for deep 
inelastic scattering processes is required. At present, it is recognized that insufficient understanding of these effects leads to approximately 
25$\%$ uncertainty in the (anti)neutrino-nucleus cross sections, contributing significantly to systematic errors. Hence, for improved precision in 
the measurement of (anti)neutrino-nucleus interaction cross sections$-$ relevant for the determination of oscillation parameters, the extraction of 
nuclear structure functions in the electroweak sector, and other related observables$-$ it is essential to achieve a better understanding of nuclear medium effects, particularly in the low-to-moderate $Q^2$ region (1-5 GeV$^2$).

In the literature, various phenomenological groups~\cite{Hirai:2007sx, Eskola:2009uj, deFlorian:2011fp, Kovarik:2012zz, Kovarik:2015cma, private} have proposed the nuclear parton distribution functions 
which may provide direct information about the nuclear modifications of the 
nucleon structure functions and the scattering cross sections. These nuclear parton distribution functions(nPDFs) are obtained by applying nuclear correction 
factors to free-nucleon PDFs while analyzing the experimental data from 
charged-lepton DIS, Drell-Yan (DY) production in proton-nucleus collisions,
and high-energy collider data from the proton-nucleus and nucleus-nucleus collisions. The global fitting groups such as those led by Hirai et al.~\cite{Hirai:2007sx}, Eskola et al.~\cite{Eskola:2009uj},
and de Florian-Sassot~\cite{deFlorian:2011fp} have included a wide range of charged-lepton and hadronic data in their analysis, often assuming that the
same nuclear correction factors apply to both the electromagnetic and the weak DIS processes. In contrast, the nCTEQ 
Collaboration~\cite{Kovarik:2012zz, private} has performed 
dedicated fits to neutrino-nucleus DIS data, extracting nuclear PDFs directly without relying solely on the constraints 
from the analysis of charged-lepton DIS data. Their analyses indicate 
potential differences between nuclear modifications in the electromagnetic and the weak structure functions, particularly in the low-$x$ region. 
This tension remains an open issue, underscoring the need for high-precision data as well as the need for developing a better theoretical understanding
of the neutrino-nucleus DIS process~\cite{SajjadAthar:2020nvy}. 

In the present work, we have studied the effects of various nuclear medium effects on the (anti)neutrino-argon DIS cross sections, 
such as the Fermi motion, the binding energy and the nucleon correlations. The nuclear medium effects are incorporated through the use of spectral
function of the nucleon in the nuclear medium~\cite{Marco:1995vb, FernandezdeCordoba:1991wf}. The effect of mesonic contribution 
has been included which is found to be significant in the low and intermediate region of $x$~\cite{Marco:1995vb}. We have also included the effect of 
shadowing and antishadowing corrections following the works of Kulagin and Petti~\cite{Kulagin:2004ie}. Moreover, we have discussed the 
effect of the center of mass energy cut on the differential scattering cross sections. 
The nucleon structure functions which are used as an input to evaluate the nuclear structure functions required to determine the (anti)neutrino-nucleus differential cross sections
are evaluated by using the MMHT parameterization of parton distribution functions (PDFs)~\cite{Harland-Lang:2014zoa} up to next-to-next-to-leading order (NNLO) in the four
flavor($u,~d,~s,$ and $c$) MSbar scheme following Ref.~\cite{Vermaseren:2005qc, Moch:2004xu, Moch:2008fj}, and the target mass correction (TMC) is 
included following the works of Schienbein et al.~\cite{Schienbein:2007gr}. 
 
 In section~\ref{sec_formalism}, we present the formalism in brief for describing the charged current induced (anti)neutrino-nucleon 
 DIS process, followed by (anti)neutrino-nucleus DIS process. In 
 section~\ref{sec_results}, the numerical results for the double and single differential scattering cross sections 
 are presented and discussed. In the last section~\ref{summary}, we summarize our findings.

\section{Deep inelastic scattering of (anti)neutrino from nucleon ($N$)/nuclei ($A$)}\label{sec_formalism}
\subsection{$\nu_\mu/\bar\nu_\mu-N$ DIS}
\begin{figure}[h]
 \centering\includegraphics[height=4.0 cm, width=6 cm]{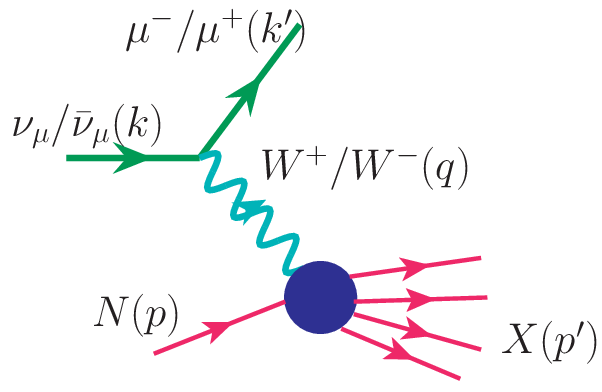}
 \caption{Feynman diagram representing the deep inelastic scattering induced by $\nu_\mu/\bar\nu_\mu-N$ interaction processes.}
 \label{fey0}
\end{figure}

For the charged current (anti)neutrino deep inelastic scattering off a free nucleon target (depicted in Fig.~\ref{fey0}),
\begin{eqnarray}\label{reaction}
 \nu_\mu(k) / \bar\nu_\mu(k) + N(p) \rightarrow \mu^-(k') / \mu^+(k') + X(p')\;,
\end{eqnarray}
the invariant matrix element is given by
\begin{equation}\label{matrix}
 -i{\cal M}=\frac{iG_F}{\sqrt{2}}\;l_\mu \;\left(\frac{M_W^2}{q^2-M_W^2} \right)\;\langle X|J^\mu|N\rangle\;,
\end{equation}
where $k$($k'$) are the four momenta of incoming(outgoing) lepton, $p$ and $p'$ are the four momenta of the target nucleon and the jet of hadrons produced 
in the final state, respectively. $G_F$ is the Fermi coupling constant, $M_W$ is the mass of $W$ boson, and $q^2=(k-k')^2=-Q^2$ ($Q^2 \ge 0$) is the four momentum transfer square. 
$l_\mu$ is the leptonic current and $\langle X|J^\mu|N\rangle$ is the 
hadronic current for the (anti)neutrino induced reaction. The differential scattering cross section 
corresponding to the reaction given in Eq.~\ref{reaction} in the lab frame is expressed as~\cite{Zaidi:2019asc}:
\begin{equation}
\label{eq:w1w2w3}
\frac{ d^2\sigma_N  }{ dx dy } =  \frac{y M_N}{\pi }~\frac{E}{E'}~\frac{|{\bf k^\prime}|}{|{ \bf k}|}\;   \overline{\sum} \sum |{\cal M}|^2 \;,
\end{equation}
where $x=\frac{Q^2}{2 M_N \nu}$ is the Bjorken scaling variable, $y=\frac{p . q}{p.k}(=\frac{\nu}{E}~\rm{in~the~lab~frame})$ is the inelasticity,
$\nu=E-E'$ is the energy transfer, $M_N$ is the nucleon mass, $E(E')$ is the energy of the incoming(outgoing) lepton. The invariant
matrix element square $\overline{\sum} \sum |{\cal M}|^2$ is given in terms of the leptonic ($L_{\mu\nu} $) and hadronic ($ W^{\mu\nu}_N$) tensors as
\begin{equation}\label{amp_wk}
 \overline{\sum} \sum |{\cal M}|^2 = \frac{G_F^2}{2}~\left(\frac{M_W^2}{Q^2+M_W^2}\right)^2 ~L_{\mu\nu} ~W^{\mu\nu}_N.
\end{equation}
 The leptonic tensor $L_{\mu \nu} $ is given by 
\begin{eqnarray}\label{lep_weak}
L_{\mu \nu} &=&8(\underbrace{k_{\mu}k'_{\nu}+k_{\nu}k'_{\mu}
-k.k^\prime g_{\mu \nu} }_{symmetric} \pm \underbrace{i \epsilon_{\mu \nu \rho \sigma} k^{\rho} 
k'^{\sigma}}_{antisymmetric})\,,
\end{eqnarray}
which consists of both the symmetric and antisymmetric terms arising due to the contribution from vector and the axial-vector components. In the 
antisymmetric term +ve sign corresponds to antineutrino case and -ve sign for the neutrino case.

The nucleon hadronic tensor $W_{N}^{\mu \nu}$ is expressed in terms of the nucleon structure functions $W_{iN}(\nu,Q^2);~(i=1-5)$ as~\cite{Zaidi:2019asc}:
\begin{eqnarray}\label{had_weak_red}
W_{N}^{\mu \nu} 
&=&\left( \frac{q^{\mu} q^{\nu}}{q^2} - g^{\mu \nu} \right) \;
W_{1N} (\nu,Q^2)
+ \frac{W_{2N} (\nu,Q^2)}{M_N^2}\left( p^{\mu} - \frac{p . q}{q^2} \; q^{\mu} \right)
\left( p^{\nu} - \frac{p . q}{q^2} \; q^{\nu} \right)
 \nonumber\\
&&-\frac{i}{2M_N^2} \epsilon^{\mu \nu \rho \sigma} p_{ \rho} q_{\sigma}~
W_{3N} (\nu,Q^2)+\frac{W_{4N} (\nu, Q^2)}{M_N^2} q^{\mu} q^{\nu}+\frac{W_{5N} (\nu, Q^2)}{M_N^2} (p^{\mu} q^{\nu} + q^{\mu} p^{\nu}).
\end{eqnarray}
In the massless lepton limit ($m_l \to 0$), when the leptonic tensor (Eq.\ref{lep_weak}) is contracted with the hadronic tensor (Eq.~\ref{had_weak_red}), the contributions from the terms associated with 
$W_{4N} (\nu, Q^2)$ and $W_{5N} (\nu, Q^2)$ vanish. 

Generally, the cross sections (Eq.\ref{eq:w1w2w3}) are expressed in terms of the dimensionless nucleon structure functions $F_{iN} (x,Q^2)$; ($i=1-3$), which
are related to the nucleon structure functions $W_{iN} (\nu,Q^2)$; ($i=1-3$) as:
\begin{eqnarray}
F_{1N} (x,Q^2)&=& M_N W_{1N} (\nu, Q^2),\\
F_{2N} (x,Q^2)&=&\nu W_{2N} (\nu, Q^2),\\
F_{3N} (x,Q^2)&=&\nu W_{3N} (\nu, Q^2).
\end{eqnarray}
Using these dimensionless nucleon structure functions, the differential scattering cross section is given by~\cite{Zaidi:2019asc}:
\begin{eqnarray}\label{d2sigdxdy_weak1}
\frac { d^2\sigma_N  }{ dx dy }&=& \frac{G_F^2 M_N E}{\pi} \left( \frac{M_W^2}{M_W^2+Q^2}\right)^2 
\left[x y^2 F_{1N} (x,Q^2) + \left(1-y-\frac{M_N x y}{ 2 E} \right) F_{2N} (x,Q^2)\pm x y \left(1-\frac{y}{2} \right)F_{3N} (x,Q^2)\right]\;.
\end{eqnarray} 
In the quark parton model, the dimensionless nucleon structure functions $(F_{2N} (x)$ and $xF_{3N} (x))$ are in turn written in terms of 
the parton distribution functions at the leading order of perturbative QCD as~\cite{halzen_martin}:
\begin{eqnarray}
F_{2N} (x)  &=& \sum_{i} x [q_i(x) +\bar q_i(x)] \;,\\
x F_{3N} (x)&=&  \sum_i x [q_i(x) -\bar q_i(x)],
\end{eqnarray} 
where $i$ runs for the different flavors of (anti)quark, and $q_i(x)(\bar q_i(x))$ corresponds to the probability density of 
finding a quark(antiquark) with a momentum fraction $x$. Assuming massless and collinear spin-$\frac{1}{2}$ partons, the structure functions in 
charged current (anti)neutrino-nucleon DIS satisfy the Callan-Gross relation~\cite{Callan:1969uq}:
\begin{equation}\label{cg}
F_{1N} (x)  = \frac{F_{2N} (x)}{2x}.
\end{equation}
Therefore, the differential scattering cross sections (Eq.\ref{d2sigdxdy_weak1}) may be written as: 
\begin{eqnarray}\label{d2sigdxdy_weak1a}
\frac { d^2\sigma_N  }{ dx dy }&=& \frac{G_F^2 M_N E}{\pi} \left( \frac{M_W^2}{M_W^2+Q^2}\right)^2 
\left[\left(1-y-\frac{y^2}{2}-\frac{M_N x y}{ 2 E} \right) F_{2N} (x)\pm x y \left(1-\frac{y}{2} \right)F_{3N} (x)\right]\;.
\end{eqnarray} 
However, in realistic (anti)neutrino-induced processes, the Callan-Gross relation 
is modified due to higher-order QCD effects, target mass corrections, and higher-twist contributions associated with multi-parton 
correlations at low to moderate values of $Q^2$, and is given at the leading order(LO) by:
\begin{equation}\label{cgm}
 F_{1N}(x,Q^2)=\frac{1}{2x}\;\Big[(1+r^2) \; F_{2N}(x,Q^2)-F_{LN}(x,Q^2)\Big],\;\;r^2=\frac{4 M^2 x^2}{Q^2},
\end{equation}
where $F_{LN}(x,Q^2)$ is the longitudinal nucleon structure function which is evaluated following the work of Moch et al.~\cite{Moch:2004xu}.
In our study, we incorporate 
higher-order QCD effects, target mass corrections, and higher-twist contributions in evaluating the nucleon structure functions.
\begin{figure}
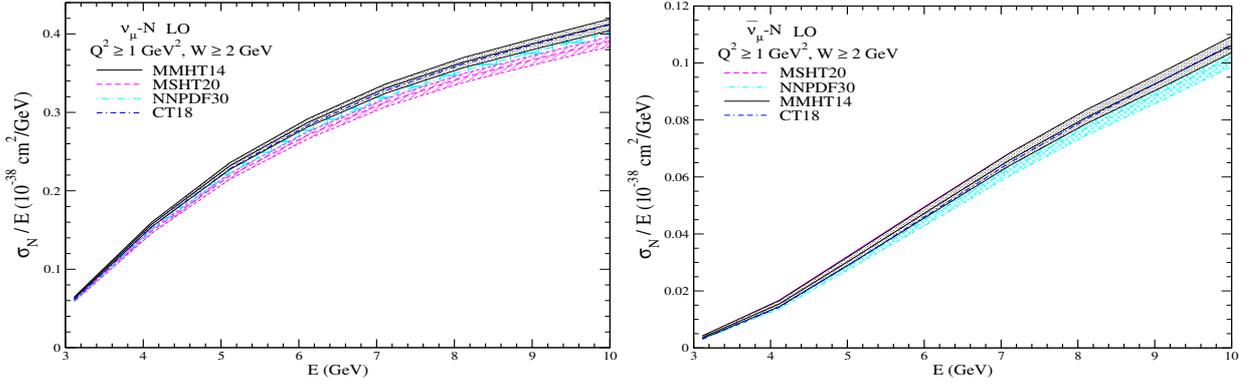

\begin{center}
 \includegraphics[height= 5 cm , width= 0.45\textwidth]{sigma_free_lo_nu.eps}
  \includegraphics[height= 5 cm , width= 0.45\textwidth]{sigma_free_lo_nubar.eps}
\end{center}
\caption{$\frac{\sigma_N}{E}$ as a function of $E$ for the charged current $\nu_\mu-N$ (left panel) and $\bar\nu_\mu-N$ (right panel) DIS processes. The
results are computed at LO using different nucleonic PDFs~\cite{Harland-Lang:2014zoa, NNPDF:2014otw, Bailey:2020ooq, Yan:2022pzl}. The 
associated PDFs uncertainties, shown as bands, correspond to symmetric 68\% confidence level about the central value. A cut of $W\ge2$ GeV is applied in the 
calculations of total scattering cross sections.}
\label{fig0a}
\end{figure}
 In the literature, different parameterizations are available for the parton density distribution
functions, namely, GRV~\cite{Gluck:2007ck}, CT10~\cite{Lai:2009ne}, CTEQ18~\cite{Yan:2022pzl}, 
MMHT14~\cite{Harland-Lang:2014zoa}, MSHT20~\cite{Bailey:2020ooq}, ABMP16~\cite{Alekhin:2018pai}, NNPDF~\cite{NNPDF:2014otw}, etc., to evaluate the nucleon 
structure functions. 

In the present work, we study the dependence of (anti)neutrino-nucleon scattering cross sections on the choice of different PDFs parameterizations in the
few-GeV energy region, which are currently used in many studies. We have evaluated the results of total scattering cross sections ($\sigma_N$) for the neutrino
as well as the antineutrino induced charged current DIS processes off free nucleon target 
at the leading order (LO) of perturbative QCD by using some of the recent PDFs parameterizations
like CT18~\cite{Yan:2022pzl}, NNPDF~\cite{NNPDF:2014otw}, MSHT20~\cite{Bailey:2020ooq} and MMHT14~\cite{Harland-Lang:2014zoa} as shown in Fig.~\ref{fig0a}. We 
have obtained the numerical results by incorporating the uncertainty in the PDFs at the 68\% confidence level which is shown by the symmetric band around the 
central value. It may be noticed from the figure that numerical results for the total scattering cross sections using these PDFs parameterizations 
overlap among themselves within the uncertainty bands. However, the results corresponding to their central values have some variation which is about 5\%-6\%
for neutrino and 2\%-4\% for antineutrino induced processes over the considered energy region. This uncertainty in the total scattering cross 
section for $\nu_\mu/\bar\nu_\mu-N$, arising from PDFs dependence, propagates into the evaluation of scattering cross sections for $\nu_\mu/\bar\nu_\mu-A$, 
where additional uncertainty arises due to the limited understanding of nuclear medium effects. In the next subsection, we briefly discuss the 
formalism for the (anti)neutrino induced DIS processes on nuclear targets and the various nuclear medium effects such as 
Fermi motion, binding energy, nucleon correlations, mesonic contributions and (anti)shadowing effect 
which have been taken into account using a microscopic many-body field theoretical approach.

\subsection{$\nu_\mu/\bar\nu_\mu-A$ DIS}
\begin{figure}[h]
 \centering\includegraphics[height=4.0 cm, width=6 cm]{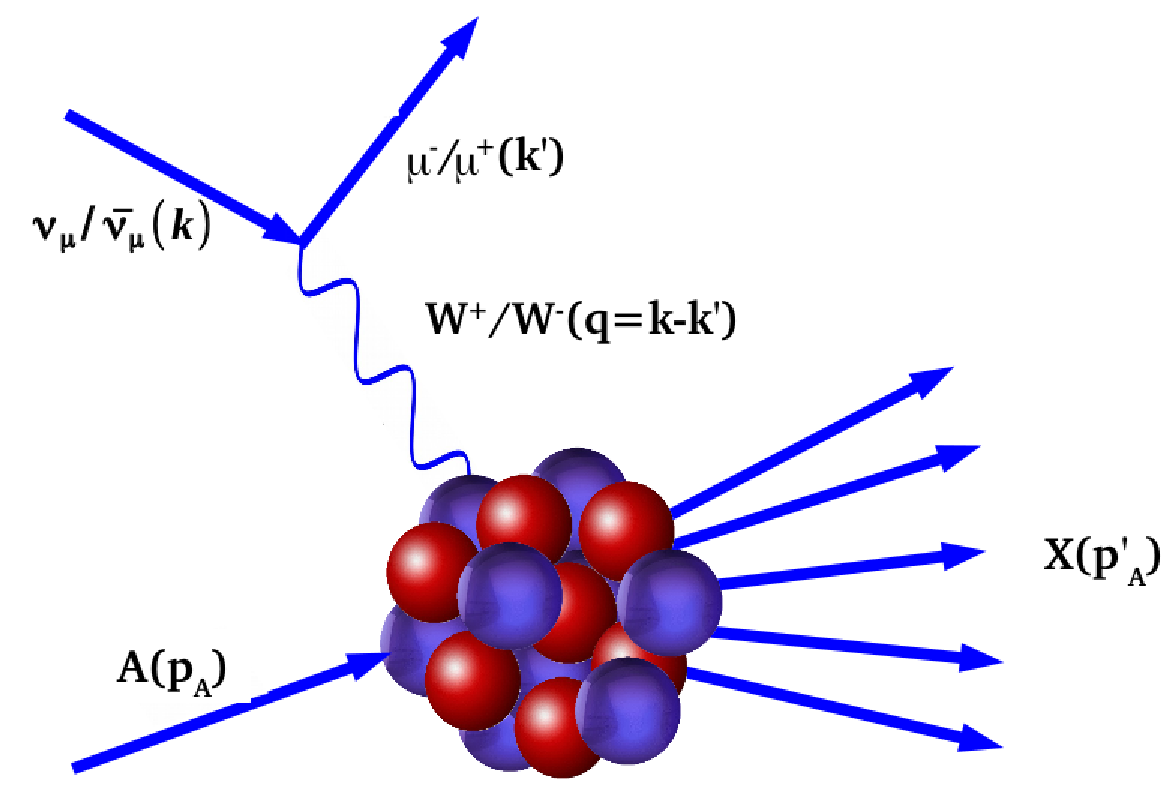}
 \caption{Feynman diagram showing the deep inelastic scattering induced by $\nu_\mu/\bar\nu_\mu-A$ interaction processes.}
 \label{fey}
\end{figure}

The basic reactions for the (anti)neutrino induced charged current DIS off a nuclear target, depicted in Fig.~\ref{fey} (with the four momenta of the particles mentioned in the corresponding
 parenthesis) are given by
 $$\nu_\mu(k) / \bar\nu_\mu(k) + A(p_A) \rightarrow l^-(k') / l^+(k') + X(p'_A),$$
for which the differential scattering cross sections are generally expressed in terms of the 
leptonic and nuclear hadronic tensors as~\cite{Zaidi:2019mfd}:
\begin{eqnarray}\label{d2sig_wi}
\frac{ d^2\sigma}{ dx dy }&=& \frac{G_F^2 \; M_N\; y}{ 2\pi}\;\frac{E}{E'} \frac{|{\bf k^\prime}|}{|{ \bf k}|} \left(\frac{M_W^2}{ M_W^2+Q^2 }\right)^2
L_{\mu\nu} \; W^{\mu\nu}_A\;.
\end{eqnarray}
 The leptonic tensor $L_{\mu\nu} $ is already defined in Eq.~\ref{lep_weak}. 
For $\nu_\mu(\bar\nu_\mu)-A$ DIS process, the nuclear hadronic tensor $W^{\mu\nu}_A$ 
is written in terms of the weak nuclear structure functions $W_{iA} (\nu,Q^2)$ ($i=1-5$) as~\cite{Zaidi:2019asc}:
\begin{eqnarray}
 \label{nuc_had_weak}
 W^{\mu\nu}_A &=& \left( \frac{q^{\mu} q^{\nu}}{q^2} - g^{\mu \nu} \right) \;
W_{1A} (\nu,Q^2)
+ \frac{W_{2A} (\nu,Q^2)}{M_A^2}\left( p_A^{\mu} - \frac{p_A . q}{q^2} \; q^{\mu} \right)
\left( p_A^{\nu} - \frac{p_A . q}{q^2} \; q^{\nu} \right)
 \nonumber\\
&&-\frac{i}{2M_A^2} \epsilon^{\mu \nu \rho \sigma} p_{A \rho} q_{\sigma}~
W_{3A} (\nu,Q^2)+\frac{W_{4A} (\nu, Q^2)}{M_A^2} q^{\mu} q^{\nu}+\frac{W_{5A} (\nu, Q^2)}{M_A^2} (p^{\mu}_A q^{\nu} + q^{\mu} p^{\nu}_A),
\end{eqnarray}
with $M_A$ as the nuclear target mass. In the massless lepton limit, terms associated with 
$W_{4A} (\nu, Q^2)$ and $W_{5A} (\nu, Q^2)$ vanish when contracted with the leptonic tensor. 

These nuclear structure functions $W_{iA} (\nu,Q^2)$; ($i=1-3$) are generally expressed in terms of the dimensionless nuclear structure 
functions $F_{iA} (x,Q^2)$; ($i=1-3$) defined as follows:
\begin{eqnarray} 
\label{eq:f1w1nua}
 F_{1A} (x,Q^2) &=& M_A ~W_{1A} (\nu,Q^2)   \ , \\
\label{eq:f2w2nua}
 F_{2A} (x,Q^2) &=& \nu ~W_{2A} (\nu,Q^2)  \ , \\
\label{eq:f3w3nua}
  F_{3A} (x,Q^2) &=& \nu ~W_{3A} (\nu,Q^2)  \ .
\end{eqnarray}
Using the above relations and Eqs.~\ref{nuc_had_weak} and \ref{lep_weak} in Eq.~\ref{d2sig_wi}, the differential cross sections is written in terms of the 
first three dimensionless nuclear structure functions $F_{iA} (x,Q^2)$; ($i=1-3$) as~\cite{Zaidi:2019asc}:
\begin{eqnarray}\label{d2sigdxdy_weak2}
\frac { d^2\sigma  }{ dx dy }&=& \frac{G_F^2 M_N E}{\pi} \left( \frac{M_W^2}{M_W^2+Q^2}\right)^2 
\left[x y^2 F_{1A} (x,Q^2) + \left(1-y-\frac{M_N x y}{ 2 E} \right) F_{2A} (x,Q^2)\pm x y \left(1-\frac{y}{2} \right)F_{3A} (x,Q^2)\right]\;.
\end{eqnarray}

As the interaction takes place with a nuclear target, the nucleons bound inside it are off-shell and continuously moving with non-zero Fermi momentum. 
For the present numerical calculations, we choose the three momentum transfer ${\bf q}$ 
to be along the $z$ axis, and therefore $q^\mu=(q^0,{\bf q})=(q^0,0,0,q^z)$, where $q^0$ is the energy transferred to the nuclear target. The Bjorken variables for 
the nuclear target($x_A$) and the bound nucleon($x_N$) are defined as
\begin{eqnarray}
x_A&=&\frac{Q^2}{2 p_A \cdot q}= \frac{Q^2}{2 M_{A}  q^0}= \frac{Q^2}{2 A~M_N q^0}\;, \hspace{6 mm}
x_N = \frac{Q^2}{2 p \cdot q} = \frac{Q^2}{2 (p^0 q^0 - p^z q^z)}.
\end{eqnarray}
The nucleons bound inside a nucleus may also interact among themselves via strong interaction which we incorporate by taking
the nucleon-nucleon correlations into account using a microscopic field theoretical approach. 
% We also ensure the binding energy for a given nucleus. 
The effect of Fermi motion, binding energy and nucleon correlations are included through the relativistic nucleon spectral function ($S_h$) which is 
 obtained by using the Lehmann's representation for the relativistic nucleon propagator. 
%  We use the technique of nuclear many body theory to calculate the dressed nucleon propagator in an interacting Fermi sea in the nuclear matter. 
The dressed relativistic nucleon propagator in an interacting Fermi sea in the infinite nuclear matter 
is calculated by using the technique of many body theory~\cite{FernandezdeCordoba:1991wf}.
 To obtain the results for a finite nucleus the local density approximation (LDA) is then applied in which
 the Fermi momentum of an interacting nucleon is not considered to be a constant quantity but the function of 
 position coordinate ($r$)~\cite{FernandezdeCordoba:1991wf}. 
 Following our earlier work~\cite{Zaidi:2019asc}, the expressions of weak nuclear structure functions which are used to evaluate the differential scattering
 cross sections are given by:
\begin{eqnarray}
\label{conv_WA1_wk_iso}
F_{1A,N} (x_A, Q^2) &=& 4AM_N \int \, d^3 r \, \int \frac{d^3 p}{(2 \pi)^3} \, 
\frac{M_N}{E_N ({\bf p})} \, \int^{\mu}_{- \infty} d p^0~ S_h(p^0, {\bf p}, \rho(r))~\nonumber\\
&\times& \left[\frac{F_{1N} (x_N, Q^2)}{M_N} + \left(\frac{p^x}{M_N}\right)^2 \frac{F_{2N} (x_N, Q^2)}{\nu_N}\right],\\
  \label{had_ten151_wk_iso}
F_{2A,N} (x_A,Q^2)  &=&  4 \int \, d^3 r \, \int \frac{d^3 p}{(2 \pi)^3} \, 
\frac{M_N}{E_N ({\bf p})} \, \int^{\mu}_{- \infty} d p^0 ~S_h (p^0, {\bf p}, \rho(r))\left(\frac{M_N}{p^0~-~p^z~\gamma}\right) \times F_{2 N} (x_N,Q^2) \nonumber \\
&\times& \left[\frac{Q^2}{(q^z)^2}\left( \frac{|{\bf p}|^2~-~(p^{z})^2}{2M_N^2}\right) +  \frac{(p^0~-~p^z~\gamma)^2}{M_N^2} \left(\frac{p^z~Q^2}{(p^0~-~p^z~\gamma) q^0 q^z}~+~1\right)^2\right]~\\
% &\times&\left(\frac{M_N}{p^0~-~p^z~\gamma}\right) \times F_{2 N} (x_N,Q^2),       \\
\label{f3a_weak}
 F_{3A,N} (x_A,Q^2) &=& 4 A  \int \, d^3 r \, \int \frac{d^3 p}{(2 \pi)^3} \, 
\frac{M_N}{E_N ({\bf p})} \, \int^{\mu}_{- \infty} d p^0 S_h (p^0, {\bf p}, \rho(r))\times \frac{q^0}{q^z} \left({p^0 q^z - p^z q^0  \over p \cdot q} \right)F_{3N} (x_N,Q^2),
\end{eqnarray}
where $\nu_N=\frac{p\cdot q}{M_N}=\frac{p^0 q^0 - p^z q^z}{M_N}$, $\gamma=\sqrt{1+\frac{4 M_N^2 x^2}{Q^2}}$ and $E_N$ is the relativistic energy
of an on-shell nucleon.
It is notable that $F_{1A,N} (x_A, Q^2)$ has been evaluated without using the Callan-Gross relation at the nuclear level~\cite{Zaidi:2019asc}.
The results obtained by using Eqs.~\ref{conv_WA1_wk_iso}, \ref{had_ten151_wk_iso} and \ref{f3a_weak} 
are labeled as the results with the spectral function(SF) only. 
 
 In the case of bound nucleons which interact among themselves via the exchange of virtual mesons 
 such as $\pi,~\rho,$ etc., there is a finite non-zero probability that intermediate vector boson may interact with these mesons instead of the nucleon. 
 Therefore, we also incorporate the mesonic contribution by using the many-body field theoretical approach similar to the 
 case of bound nucleons following Ref.~\cite{Marco:1995vb}. The expressions for the mesonic structure functions $F_{1A, i} (x,Q^2)$ and
 $F_{2A, i} (x,Q^2)$ for pions and rho meson are given by~\cite{Zaidi:2019asc}
\begin{eqnarray} 
\label{F2rho_wk}
F_{1A, i} (x,Q^2) &=& - 6\times a \times A M_N \int  d^3 r   \int  \frac{d^4 p}{(2 \pi)^4} \theta (p^0) ~\delta I m D_i (p) \;2m_i\nonumber\\
&\times& ~\left[\frac{F_{1i} (x_i)}{m_i}~+~\frac{{|{\bf p}|^2~-~(p^{z})^2}}{2(p^0~q^0~-~p^z~q^z)}
\frac{F_{2i} (x_i)}{m_i}\right],\\
 \label{F2rho1_wk}
F_{2A, i} (x,Q^2)  &=& - 6\times a \int \, d^3 r \, \int \frac{d^4 p}{(2 \pi)^4} \, 
        \theta (p^0) ~\delta I m D_i (p) \;2m_i\left(\frac{m_i}{p^0~-~p^z~\gamma}\right) ~F_{2i} (x_i)\nonumber \\
&\times&\left[\frac{Q^2}{(q^z)^2}\left( \frac{|{\bf p}|^2~-~(p^{z})^2}{2m_i^2}\right)  
+  \frac{(p^0~-~p^z~\gamma)^2}{m_i^2} \left(\frac{p^z~Q^2}{(p^0~-~p^z~\gamma) q^0 q^z}~+~1\right)^2\right],~\nonumber\\
% &\times& \left(\frac{m_i}{p^0~-~p^z~\gamma}\right) ~F_{2i} (x_i),
\end{eqnarray}
where subscript $i=\pi/\rho$, $m_i$ is the mass of pion/rho meson, $x_i=\frac{Q^2}{-2p \cdot q}$ with an extra minus sign due to the direction
of mesonic four momenta $p$, and $a=1$ for pion and $a=2$ for rho meson~\cite{Marco:1995vb}. Notice that the $F_{2A, \rho} (x,Q^2)$ has 
an extra factor of two compared to pionic contribution because it has two transverse polarization states~\cite{Baym:1975vb}.
 For the numerical calculations, we use the GRV PDFs parameterization~\cite{Gluck:1991ey} for both the 
pion and rho meson. Recall that parity violating nuclear structure function $F_{3A} (x_A, Q^2)$ has no 
mesonic contribution as it depends mainly on the valence quarks distribution which average to zero 
when the three charge states of pion and rho mesons are taken into consideration. 
 
 Besides the mesonic contributions, the shadowing and antishadowing effects, which dominate in the low $x$ region, have also been taken into account in the present numerical calculations, following the work of Kulagin and Petti~\cite{Kulagin:2004ie} based on the Glauber-Gribov multiple scattering theory.
The nuclear structure functions $F_{iA}^{S}(x,Q^2);~(i=1-3)$ with the shadowing effect are given by
\begin{equation}
 F_{iA}^{S}(x,Q^2) = \delta R(x,Q^2) \times F_{iN} (x,Q^2)\; ,
 \label{shdw11}
\end{equation}
where superscript $S$ corresponds to the nuclear shadowing effect and the correction factor $\delta R(x,Q^2)$ is taken from Ref.~\cite{Kulagin:2004ie}.

After taking the perturbative corrections up to NNLO~\cite{Moch:2004xu, Moch:2008fj, Vermaseren:2005qc} and nonperturbative TMC~\cite{Schienbein:2007gr} effect into account for the evaluation of free nucleon structure functions
which are being used as an input to calculate the nuclear structure functions $F_{iA} (x,Q^2);~(i=1,2,3)$, the full expression for the 
nuclear structure functions $F_{iA} (x,Q^2)$ with the nuclear medium effects is given by:
\begin{equation}\label{sf_full}
  F_{iA} (x,Q^2)= F_{iA,N} (x,Q^2) + \;F_{iA, \pi} (x,Q^2)  + F_{iA, \rho} (x,Q^2) \;+ F_{iA}^{S}(x,Q^2)\;,\;i=1-2
\end{equation}
and 
\begin{eqnarray}\label{f3_tot}
 F_{3A} (x,Q^2)= F_{3A,N} (x,Q^2) + F_{3A}^S (x,Q^2).
\end{eqnarray}
It should be noted that $F_{3A} (x,Q^2)$ has no mesonic contribution.
% The corresponding results with the full model are labeled as ``Total'' in the present work. 
% Using the present formalism, and incorporating the perturbative corrections up to NNLO as well as the nonperturbative
% target mass correction following Ref.~\cite{Zaidi:2019asc}, we have presented the results for the differential scattering 
% cross sections for $\nu_\mu$ and $\bar\nu_\mu$ induced DIS off $^{40}Ar$ nuclear target in the following section. 
% For the details of the formalism of nuclear structure functions please see Refs.~\cite{Zaidi:2019asc, Zaidi:2021iam, AtharSajjad:2022ipr}. 

\section{Results and Discussion}\label{sec_results}
We present the results for the double differential \Big($\frac{d^2\sigma }{dx dy}$\Big) and single differential \Big($\frac{d\sigma }{dx }$\Big) scattering 
cross sections for $\nu_\mu$ and $\bar\nu_\mu$ induced DIS off a free nucleon ($N$) target, as well as for the nucleons bound inside the argon nucleus. 
The (anti)neutrino-nucleon scattering cross sections are calculated using the free nucleon structure functions $F_{iN}(x,Q^2);(i=1-3)$, as given 
in Eq.~\ref{d2sigdxdy_weak1}.
In our earlier works~\cite{Zaidi:2019asc, Zaidi:2019mfd}, we have demonstrated that the numerical results obtained for the 
 structure functions at NNLO with the target mass corrections, and at NLO with the target mass corrections and higher-twist effects are in agreement 
 within a few percent. Therefore, in this work, the structure functions $F_{iN}(x,Q^2);(i=1-3)$ have been evaluated at NNLO following Refs.~\cite{Zaidi:2019asc, Vermaseren:2005qc, Moch:2004xu, Moch:2008fj},
using the MMHT14 PDFs parameterization~\cite{Harland-Lang:2014zoa} in the four-flavor MSbar scheme, incorporating the TMC effect~\cite{Schienbein:2007gr}.
The corresponding results are labeled as ``Free'' in the curves shown in various figures.

The differential scattering cross sections for $\nu_\mu(\bar\nu_\mu)$ induced deep inelastic 
scattering off an argon($^{40}Ar$) target, as given in Eq.~\ref{d2sigdxdy_weak2}, 
are obtained using the nuclear structure functions $F_{iA}(x,Q^2);(i=1-3)$ defined in Eqs.~\ref{conv_WA1_wk_iso}-\ref{f3a_weak}. These $F_{iA}(x,Q^2);(i=1-3)$,
are calculated by convoluting the free nucleon structure functions $F_{iN}(x,Q^2);(i=1-3)$ with the nucleon spectral function and integrated over the entire nuclear volume.
This procedure accounts for the effect of Fermi motion, binding energy, and nucleon correlations. The corresponding results are labeled as ``SF'' in the text.

The effects of mesonic contributions arising from pion and rho meson clouds are included using Eqs.~\ref{F2rho_wk} and \ref{F2rho1_wk}, 
following a many-body field theoretical
approach discussed in our earlier works~\cite{Zaidi:2019asc, Zaidi:2019mfd, SajjadAthar:2020nvy}. Moreover, the shadowing and antishadowing corrections 
which are present in the case of scattering from nuclear targets, are incorporated(Eq.~\ref{shdw11}) following Ref.~\cite{Kulagin:2004ie}. 
The results including the contributions from the spectral function, mesonic effects, and shadowing and 
antishadowing corrections, i.e., our full model, as defined in Eqs.~\ref{sf_full} and \ref{f3_tot}, are labeled as ``Total''. 
It may be further noted 
out that $^{40}\mathrm{Ar}$ is a non-isoscalar nuclear target; however, we have treated it as isoscalar in view of the effect of non-isoscalarity
being very small ($1\text{--}2\%$) in $^{40}Ar$~\cite{SajjadAthar:2020nvy} in the considered region of Bjorken $x$ and $y$.

The present numerical calculations are performed for $Q^2\ge 1$ GeV$^2$ at two different incoming (anti)neutrino beam energies, namely $E=4$ GeV and $E=6$ GeV, 
under the following constraints on the center of mass energy $W$:
 \begin{itemize}
  \item without applying any kinematical constraint on $W$; these results are labeled as ``No W$_{cut}$''.
  \item by applying kinematical cuts on the center of mass energy $W$; specifically $W\ge 1.7$ GeV and $W\ge 2$ GeV, to study the effect of the  $W$ cut on the differential cross sections, 
  which is relevant for understanding the transition region between nucleon resonances and DIS.
  \end{itemize}

% All the results presented here are evaluated at NNLO, including the target mass correction effects, following our
% earlier works~\cite{Zaidi:2019asc, Zaidi:2019mfd}, where we have demonstrated that the numerical results obtained for the 
% structure functions at NNLO with the target mass corrections, and at NLO with the target mass corrections and higher-twist effects are in agreement 
% within a few percent. 
% Therefore, all the present results are evaluated at NNLO including the target mass correction effects.

% For the details of the formalism of nuclear structure functions please see Refs.~\cite{Zaidi:2019asc, Zaidi:2021iam, AtharSajjad:2022ipr}. 
\begin{figure}
\begin{center}
 \includegraphics[height= 8 cm , width= 0.95\textwidth]{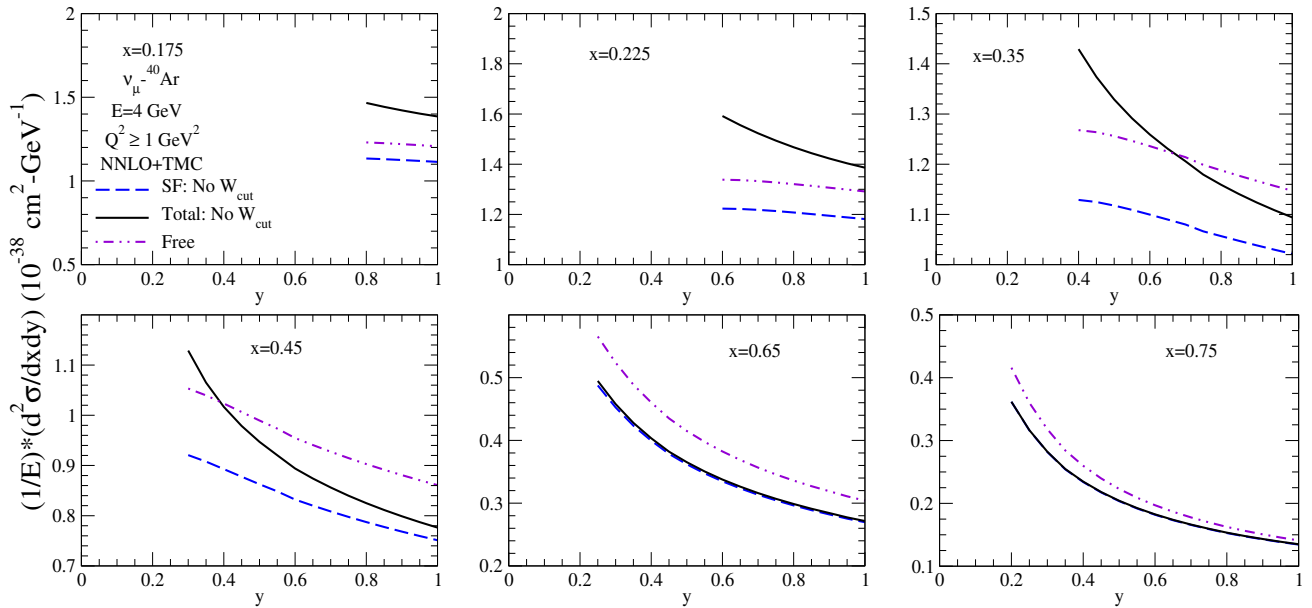}
\end{center}
\caption{$\frac{1}{E}\;\frac{d^2\sigma }{dx dy}$ as a function of $y$ at $E=4$ GeV for the charged current $\nu_\mu-^{40}Ar$ DIS process
at the different values of $x$. The results
are presented for $Q^2\ge 1$ GeV$^2$, without applying any cut on the $W$. The numerical calculations are performed
at NNLO incorporating the TMC effect using the spectral function only (dashed lines) and the full
model (solid lines). For comparison, results for a free nucleon target are also included (dash-double dotted lines).}
\label{fig0}
\end{figure}
\begin{figure}
\begin{center}
 \includegraphics[height= 8 cm , width= 0.95\textwidth]{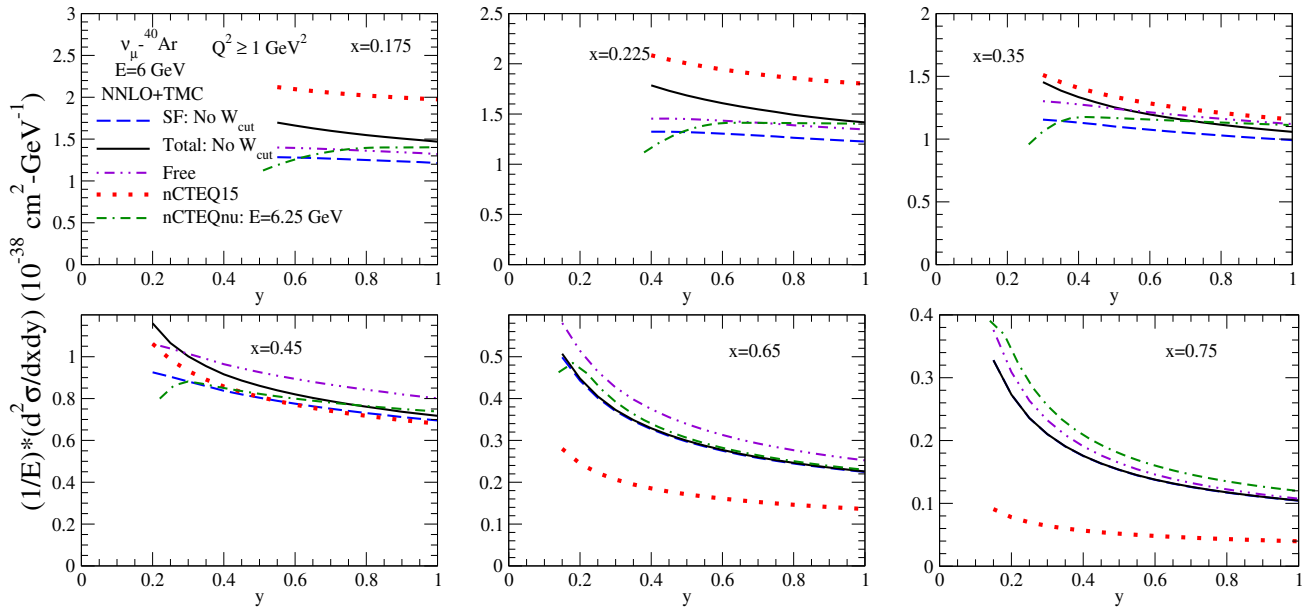}
\end{center}
\caption{$\frac{1}{E}\;\frac{d^2\sigma }{dx dy}$ as a function of $y$ for the incoming beam energy $E=6$ GeV. The line styles follow the same convention
as in Fig.~\ref{fig0}. The results are also compared with the phenomenological nuclear PDFs parameterizations:
nCTEQ15~\cite{Kovarik:2015cma} (dotted lines) and nCTEQnu~\cite{private}
(double dash-dotted lines).}
\label{fig1}
\end{figure}

 In Fig.\ref{fig0}, we present the results for $\frac{1}{E}\;\frac{d^2\sigma }{dx dy}$ as a function of $y$ for different values of Bjorken $x$ in the case of $\nu_\mu-^{40}Ar$ 
 scattering at NNLO with the TMC effect for a beam energy of 4 GeV corresponding to the spectral function only (dashed line) and 
 to the full model (solid line), without applying any cut on the center of mass energy. These results are
 compared with those for a free nucleon target (dash double-dotted line) in order to study the effects of 
 nuclear medium corrections. We find that the inclusion of nuclear structure effects through the nucleon spectral function leads to a reduction in the
 cross sections in the entire range of $x$ and $y$ considered here. For instance, the reduction is about 10-13\% at $y=0.5$ and 
 6-12\% at $y=0.8$ for $0.35\le x\le 0.75$ at $E=4$ GeV. When the mesonic contributions are included, 
 the cross section increases, while the effects of shadowing and antishadowing corrections are negligible. Consequently, the results of the full model 
 are larger than those obtained with the spectral function alone for $x\le 0.45$. For example, this enhancement from the results obtained 
 with nuclear structure effects is about 20\% at $y=0.5$ and 10\% at $y=0.8$
 for $x=0.35$. This enhancement decreases with increasing $y$, becoming about 10\% at $y=0.5$ and 5$\%$ at $y=0.8$ for $x=0.45$, and reduces further 
 with increasing $x$. Furthermore, we find that due to the inclusion of nuclear medium effects, the differential scattering cross sections using the
 full model increases about 6\% at $y=0.5$ and $x=0.35$ from the results for the free nucleon case. While at $y=0.8$ it decreases to $\sim 3\%$ 
for the same value of $x$.
 
 In Fig.~\ref{fig1}, the results for $\frac{1}{E}\;\frac{d^2\sigma }{dx dy}$ as a function of $y$ for $\nu_\mu-^{40}Ar$ are presented for $E=6$ GeV. It may 
 be noted from the figure that, with increasing energy, the 
 scattering cross sections contribute also to the lower region of inelasticity $y$. Furthermore, the nuclear medium effects due to the nuclear structure effect
 is reduced with increasing beam energy. For example, at $E=6$ GeV the suppression in the differential cross sections obtained 
 using the spectral function alone is about 11-13\% for $0.35\le x \le 0.65$ and $0.5\le y\le 0.8$, which decreases to 
 about 7\% at $y=0.5$ and 4\% at $y=0.8$ for $x=0.75$, as compared to the free nucleon case. Incorporating mesonic contributions result in an enhancement in the 
 differential cross sections which is about 27\% at $y=0.5$ and 18\% at $y=0.8$ for $x=0.225$, 
  from the results with spectral function only. While it becomes 14\% and 8\% for $x=0.35$, and 7\% and 4\% for $x=0.45$, respectively for the same values of $y$. 
  We also find that the relative increase due to mesonic contributions at $E=6$ GeV are smaller than those observed at $E=4$ GeV, with the  
reduction of about 2-5\% for $0.225 \le x \le 0.45$ and $0.5\le y\le 0.8$. The numerical results for $E=6$ GeV are 
further compared with those obtained using 
 the phenomenological nuclear PDFs parameterizations from the CTEQ group, namely nCTEQ15~\cite{Kovarik:2015cma}, which 
 includes data from charged-lepton as well as 
 neutrino induced DIS experiments, and nCTEQnu~\cite{private}, which includes data only from neutrino experiments. It 
 may be observed that the results obtained using the nCTEQ15 and nCTEQnu
 PDFs parameterizations are not mutually consistent, except in the region $0.35\le x \le 0.45$ for intermediate-to-high values of $y$.
 Our numerical results using the full model are larger than those obtained using the nCTEQnu nuclear PDFs parameterization~\cite{private}
 in the low- to intermediate- $x$ region, particularly at lower values of inelasticity;  
 however, becomes smaller with increasing $x(>0.65)$. The results based on the nCTEQ15 PDFs parameterization~\cite{Kovarik:2015cma}
 are found to be in reasonable agreement with the present calculations only in the region $0.35\le x \le 0.45$, while 
 noticeable enhancement in the low-$x(\le 0.35)$ region and reduction in the high-$x(>0.35)$ region.
\begin{figure}
\begin{center}
 \includegraphics[height= 7.8 cm , width= 0.9\textwidth]{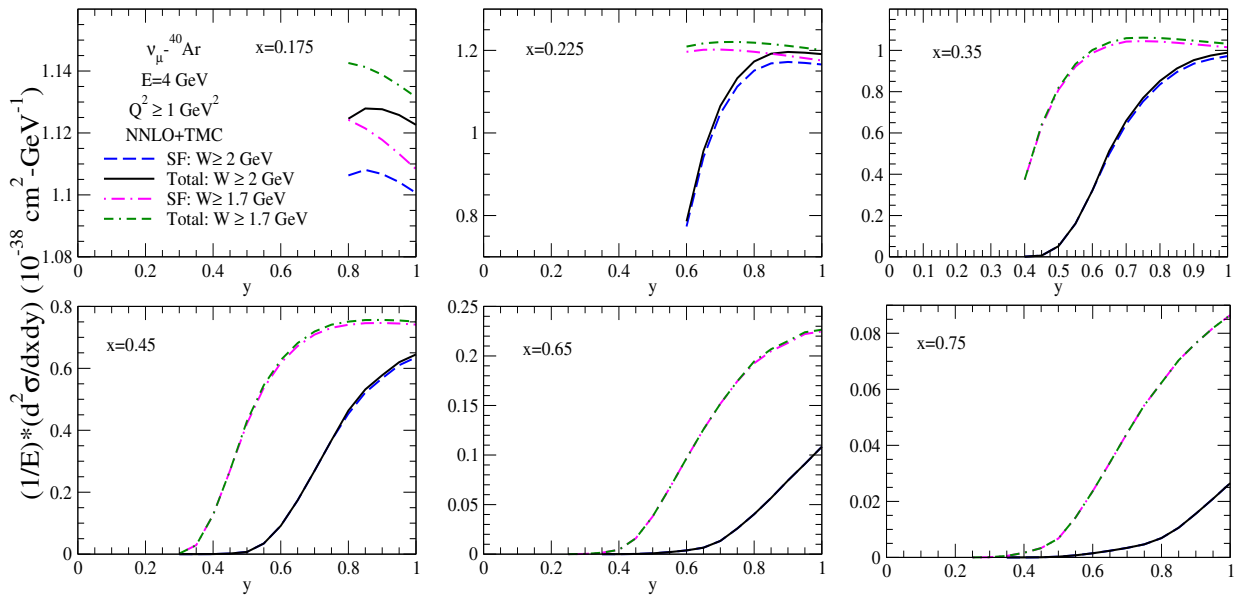}
\end{center}
\caption{$\frac{1}{E}\;\frac{d^2\sigma }{dx dy}$ as a function of $y$ at $E=4$ GeV. The line styles follow the same convention as in Fig.~\ref{fig0}, with the 
exception that different cuts on the $W$ are applied: $W\ge 1.7$ GeV and $W\ge 2$ GeV.}
\label{fig2}
\end{figure}

\begin{figure}
\begin{center}
 \includegraphics[height= 7.8 cm , width= 0.95\textwidth]{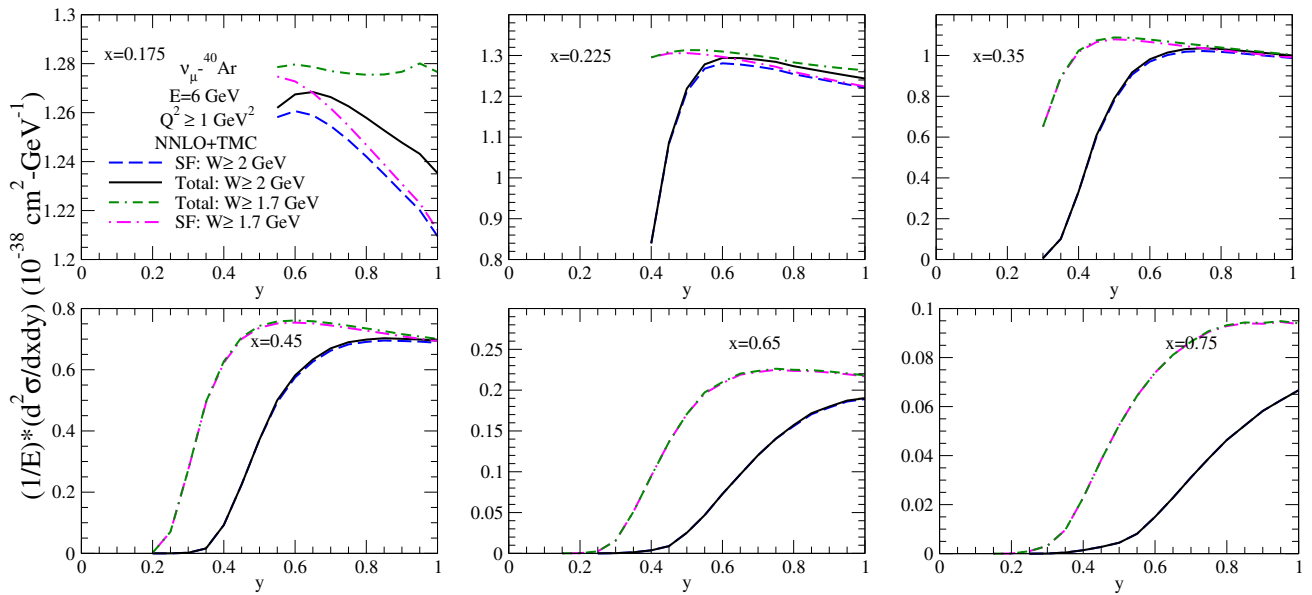}
\end{center}
\caption{$\frac{1}{E}\;\frac{d^2\sigma }{dx dy}$ as a function of $y$ at $E=6$ GeV. The line styles follow the same convention as in Fig.~\ref{fig0}, with the 
exception that different cuts on the $W$ are applied: $W\ge 1.7$ GeV and $W\ge 2$ GeV.}
\label{fig3}
\end{figure}

To illustrate the effect of the $W$ cut on the differential cross sections, we present the results
in Figs.~\ref{fig2} and \ref{fig3} for $E=4$ GeV and $E=6$ GeV, respectively. The numerical results are shown
for different choices of $W$ cuts, namely $W\ge 1.7$ GeV and $W\ge 2$ GeV, since no unique 
kinematic boundaries are defined in the literature for the transition region from nucleon resonances to deep inelastic scattering, as discussed 
earlier in the text. We find that the differential cross sections obtained with a cut on $W$ are significantly
suppressed, and this suppression becomes more pronounced with increasing $W$ cut (i.e., $W\ge 2$ GeV compared to $W\ge 1.7$ GeV).
For example, in Fig.~\ref{fig2}, when a cut of $W\ge 1.7$ GeV is applied, the suppression is about $24\%$ at $y=0.6$ and $17\%$ at $y=0.8$ for $x=0.225$. However, for $x=0.45$
the suppression becomes $88\%$ at $y=0.4$, $30\%$ at $y=0.6$ and $\sim 9\%$ at $y=0.8$. However, the results obtained with a $W\ge 2$ GeV cut are further suppressed 
by about $26\%$ at $y=0.6$ and $3\%$ at $y=0.8$ for $x=0.225$, while for $x=0.45$ the suppression is about $10\%$ at $y=0.4$, $5\%$ at $y=0.6$, and $35\%$ at $y=0.8$. 
It is important to note that the effect of the $W$ cut becomes more pronounced with increasing $x$, particularly in the 
 low- to intermediate- $y$ region. Furthermore, the application of a cut on the $W$ leads to a reduction in the mesonic cloud contribution also 
 for low to intermediate values of $x$. As expected the impact of the $W$ cut decreases with increasing neutrino beam energy. From Fig.~\ref{fig3}, it may be observed that
at $E=6$ GeV the difference between the results obtained for $W\ge 1.7$ GeV and $W\ge 2$ GeV is about 
$68\% (85\%)$ at $y=0.6$ and $20\%(38\%)\%$ at $y=0.8$ for $x=0.35(0.45)$.
% Quantitatively, the difference between the results obtained for $W\ge 1.7$ GeV and $W\ge 2$ GeV at $E=6$ GeV is 
% $68\%$ at $y=0.6$ and $20\%$ at $y=0.8$ for $x=0.35$ which becomes $85\%$ at $y=0.6$ and $38\%$ at $y=0.8$ for $x=0.45$, and 
% $96\%$ at $y=0.6$ and $78\%$ at $y=0.8$ for $x=0.65$.

  \begin{figure}
\begin{center}
 \includegraphics[height= 8 cm , width= 0.95\textwidth]{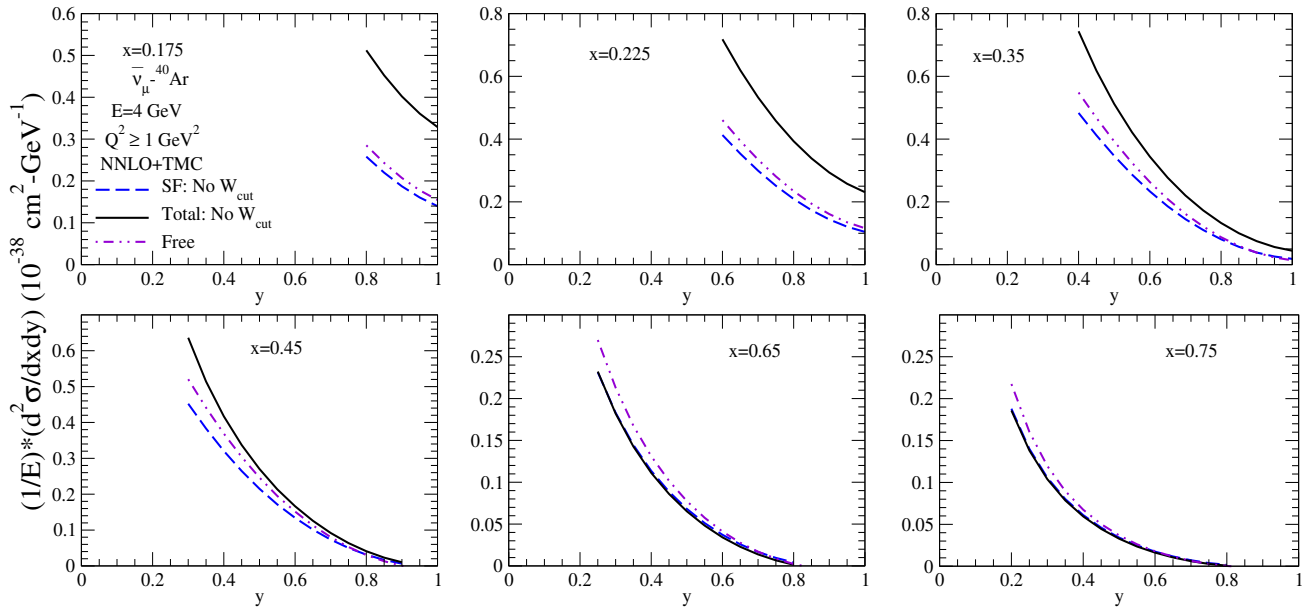}
\end{center}
\caption{$\frac{1}{E}\;\frac{d^2\sigma }{dx dy}$ as a function of $y$ at $E=4$ GeV for the charged current $\bar\nu_\mu-^{40}Ar$ DIS process
at the different values of $x$. The results
are presented for $Q^2\ge 1$ GeV$^2$, without applying any cut on the $W$. The numerical calculations are performed
at NNLO incorporating the TMC effect using the spectral function only (dashed lines) and the full
model (solid lines). For comparison, results for a free nucleon target are also included (dash-double dotted lines).}
\label{fig4}
\end{figure}

  \begin{figure}
\begin{center}
 \includegraphics[height= 8 cm , width= 0.95\textwidth]{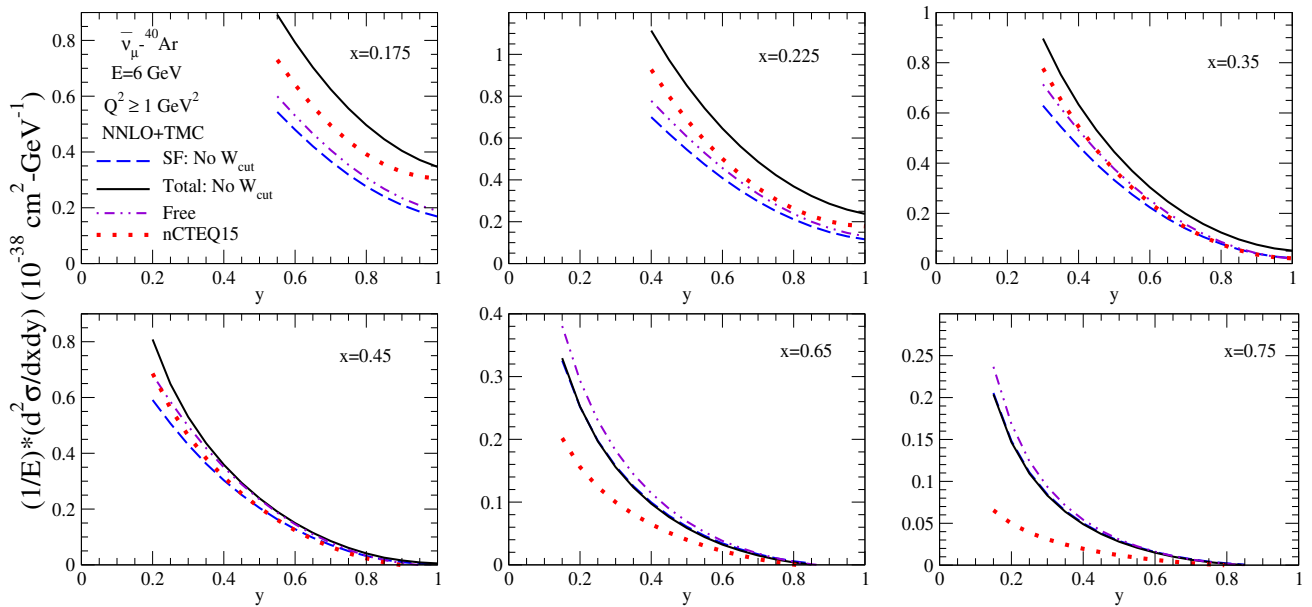}
\end{center}
  \caption{$\frac{1}{E}\;\frac{d^2\sigma }{dx dy}$ as a function of $y$ for the incoming beam energy $E=6$ GeV. The line styles follow the same convention
as in Fig.~\ref{fig4}. }
\label{fig5}
\end{figure}

In Figs.~\ref{fig4} and \ref{fig5}, the results for the $\bar\nu_\mu-^{40}Ar$ differential scattering cross sections are presented
at beam energies $E=4$ GeV and $E=6$ GeV, respectively, without any
constraint on the center of mass energy $W$. The differential cross sections $\frac{1}{E}\;\frac{d^2\sigma }{dx dy}$, for $\bar\nu_\mu-^{40}Ar$
obtained using the spectral function only (``SF'') gets suppressed due to nuclear structure effects compared to those for the free nucleon target over the entire kinematic 
region of $x$ and $y$. The qualitative dependence of differential cross sections on $x$ and $y$ is similar to that observed in 
the neutrino case(Figs.~\ref{fig0} and \ref{fig1});  however, there are 
quantitative differences. For example, the reduction
in the cross section due to effect of nuclear structure, relative to the free nucleon case, is about $13\%$ at $y=0.45$ for 
$0.35\le x \le 0.65$ at $E=4$ GeV. The inclusion of mesonic cloud contributions leads to an
enhancement in the differential cross sections for $x\le 0.45$, while the effect is very small for 
$x>0.45$. For instance, the enhancement is about $75\%$ at $y=0.65$ for $x=0.225$; $50\%$ at $y=0.45$ and $48\%$ at $y=0.65$ 
for $x=0.35$; and $27\%$ at $y=0.45$ and $24\%$ at $y=0.65$ when $x=0.45$. We find that the net effect of the nuclear medium corrections (``Total'') is to increase
the differential cross sections obtained using the full model relative to the case of free nucleon target for $x\le 0.45$, for example, this enhancement in the 
results is about 30\%(10\%) at $y=0.5$ and 51\%(35\%) at $y=0.8$ for $x=0.35(0.45)$ and $E=4$ GeV. While for $x>0.45$,
the effect of the nuclear medium corrections is to decrease the differential cross sections, for example, at $x=0.65$ and $E=4$ GeV, this decrease is about
16\% at $y=0.5$ and 38\% at $y=0.8$ from the $\bar\nu_\mu-N$ scattering cross sections.
It may be noted that the nuclear medium effects in $\bar\nu_\mu-^{40}Ar$ deep
inelastic scattering are more pronounced than in the $\nu_\mu-^{40}Ar$ case. 
The results shown in Fig.~\ref{fig5} for $E=6$ GeV
are compared with those obtained using the nCTEQ15 nuclear PDFs parameterization~\cite{Kovarik:2015cma}. We find that our numerical results are larger than 
the nCTEQ15 results in the entire range of $x$, while show reasonable agreement with them only in the intermediate $x$ region ($0.35\le x \le 0.45$). Hence,
further work is needed to better understand nuclear medium effects over the full range of Bjorken $x$.
 
\begin{figure}
\begin{center}
 \includegraphics[height= 7.0 cm , width= 0.95\textwidth]{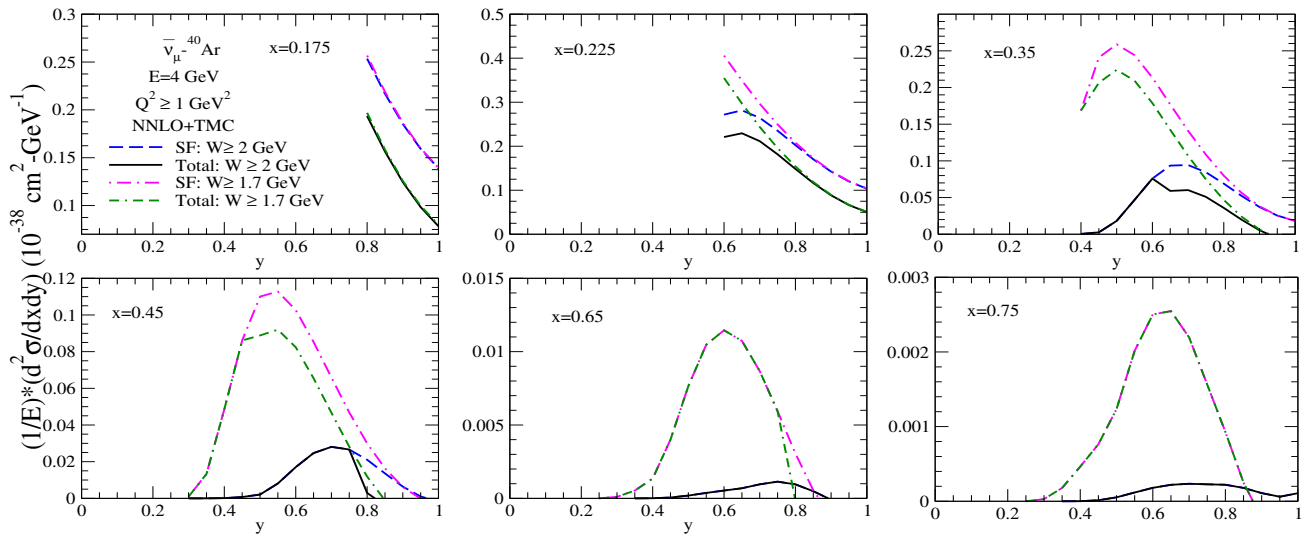}
\end{center}
\caption{$\frac{1}{E}\;\frac{d^2\sigma }{dx dy}$ as a function of $y$ at $E=4$ GeV. The line styles follow the same convention as in Fig.~\ref{fig4}, with the 
exception that different cuts on the $W$ are applied: $W\ge 1.7$ GeV and $W\ge 2$ GeV.}
% 
% \caption{The lines have the same meanings as in Fig.~\ref{fig4} except that different cuts on center mass energy ($W\ge 1.7$ GeV and $W\ge 2$ GeV)
% are applied here at $E=4$ GeV.}
\label{fig6}
\end{figure}

\begin{figure}
\begin{center}
 \includegraphics[height= 8 cm , width= 0.95\textwidth]{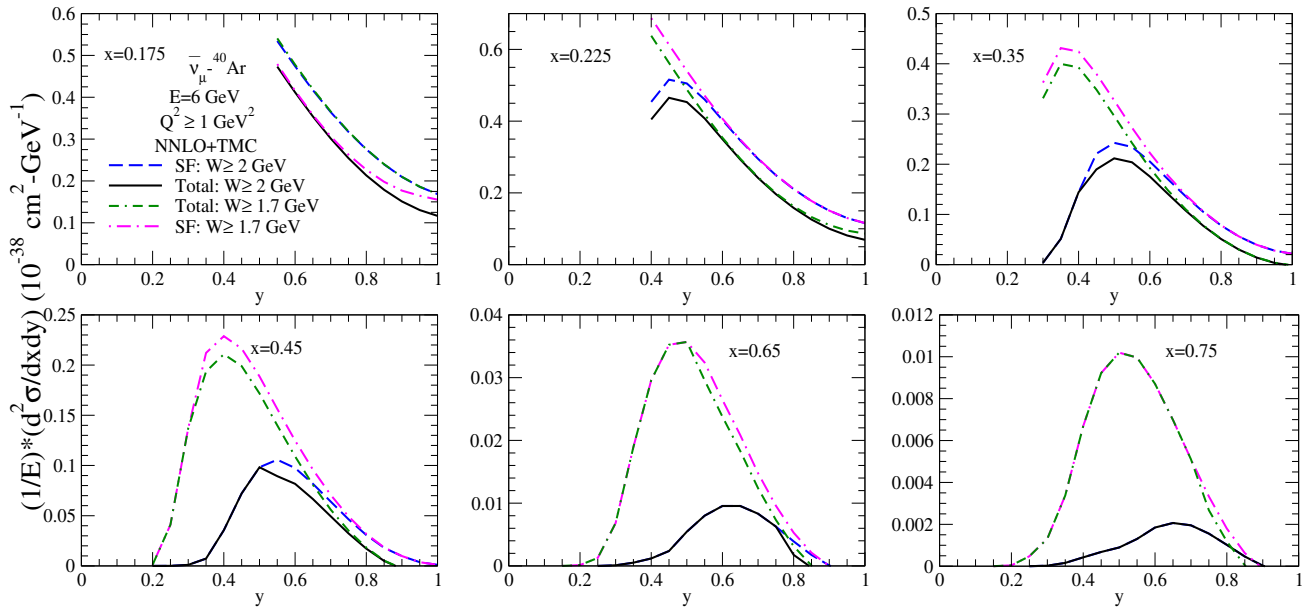}
\end{center}
\caption{$\frac{1}{E}\;\frac{d^2\sigma }{dx dy}$ as a function of $y$ at $E=6$ GeV. The line styles follow the same convention as in Fig.~\ref{fig0}, with the 
exception that different cuts on the $W$ are applied: $W\ge 1.7$ GeV and $W\ge 2$ GeV.}
\label{fig7}
\end{figure}

The numerical results for the $\bar\nu_\mu-^{40}Ar$
 differential cross sections are presented in Figs.~\ref{fig6} and \ref{fig7} by applying cuts of $W\ge 1.7$ GeV and $W\ge 2$ GeV 
 on the center of mass energy $W$ which show a suppression of the cross sections due to a cut on $W$. For example, when
 a cut of $W\ge 1.7$ GeV is applied, the suppression is about $45-50\%$ at $y=0.6$ and $56-60\%$ at $y=0.8$ 
 for $x=0.225$ and 4 GeV$\le E\le 6$ GeV, relative to the results obtained without applying any cut on $W$. It may be observed that 
 the effect of the $W$ cut is more pronounced in the antineutrino-induced processes than in the neutrino-induced ones. 
The impact of the $W$ cut on the mesonic contribution is quantitatively smaller than that observed in the $\nu_\mu$ induced process. 
 
\begin{figure}
\begin{center}
 \includegraphics[height= 7.5 cm , width= 0.95\textwidth]{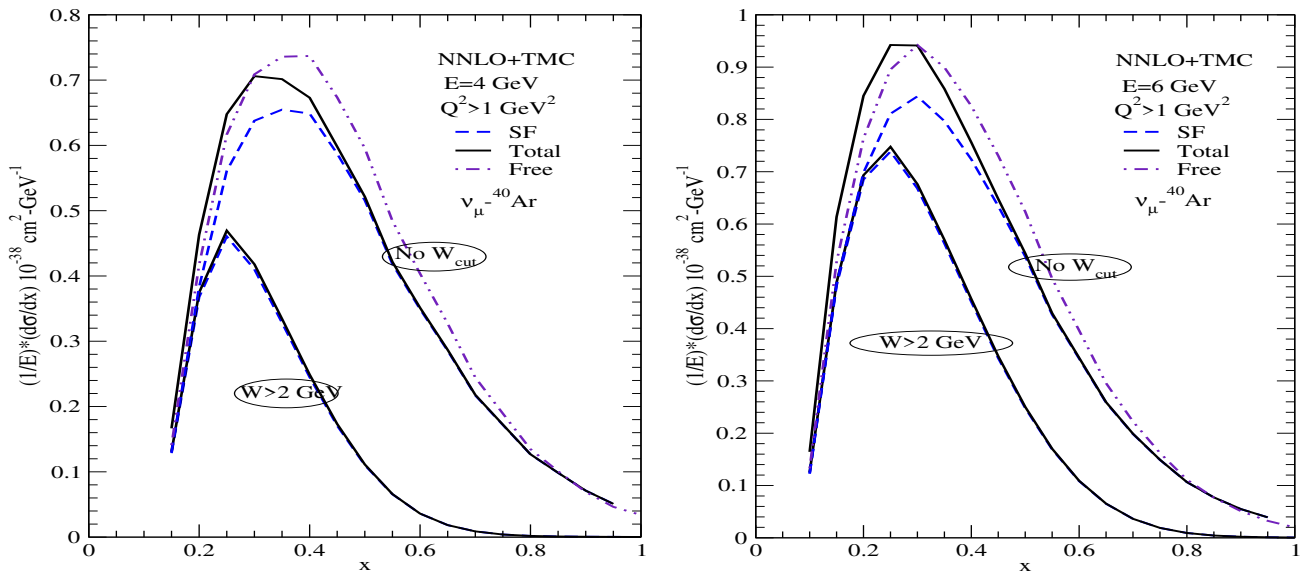}
\end{center}
\caption{$\frac{1}{E}\;\frac{d\sigma }{dx}$ as a function of $x$ for $\nu_\mu-^{40}Ar$ DIS. The upper curves correspond to results 
 without a cut on the $W$, while the lower curves include a cut $W \ge 2$ GeV. 
The calculations are performed at NNLO with the TMC effect for $E=4$ GeV (left panel) and $E=6$ GeV (right panel). Results 
are shown using the spectral function only (SF: dashed lines) and the full model (Total: solid lines). For comparison, results for a free
nucleon target are also included (dash-double dotted lines).}
\label{fig8}
\end{figure}

\begin{figure}
\begin{center}
 \includegraphics[height= 7.5 cm , width= 0.95\textwidth]{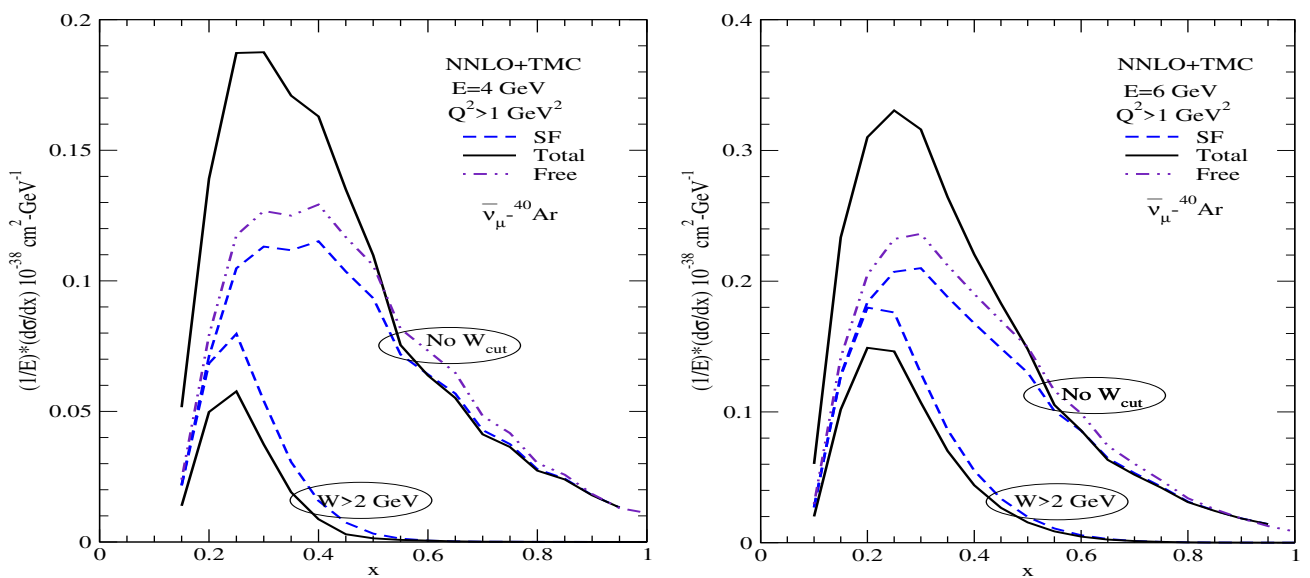}
\end{center}
\caption{$\frac{1}{E}\;\frac{d\sigma }{dx}$ as a function of $x$ for $\bar\nu_\mu-^{40}Ar$ DIS. The upper curves correspond to results 
 without a cut on the $W$, while the lower curves include a cut $W \ge 2$ GeV. 
The calculations are performed at NNLO with the TMC effect for $E=4$ GeV (left panel) and $E=6$ GeV (right panel). Results 
are shown using the spectral function only (SF: dashed lines) and the full model (Total: solid lines). For comparison, results for a free 
nucleon target are also included (dash-double dotted lines).}
\label{fig9}
\end{figure}

In Figs.~\ref{fig8} and \ref{fig9}, we present the numerical results for $\frac{1}{E}\;\frac{d\sigma }{dx}$, using Eq.~\ref{d2sigdxdy_weak1}, obtained by 
integrating $\frac{1}{E}\;\frac{d^2\sigma }{dx\;dy}$ over the entire range of inelasticity $y$, i.e., $[0,1]$.
% To show the dependence of the differential cross section in 1-D on the nuclear medium effects as well as on the choice of center of mass energy $W$, we
% perform the integration over inelasticity $y$ lying in the range of $[0,1]$ using Eq.~\ref{d2sigdxdy_weak1}, and obtain the numerical results 
% for $\frac{1}{E}\;\frac{d\sigma }{dx}$ as a function of $x$ for both the $\nu_\mu-$ and $\bar\nu_\mu-$ induced deep inelastic reactions off argon target.
These results are shown in Fig.~\ref{fig8} for $\nu_\mu$, and in Fig.~\ref{fig9} for $\bar\nu_\mu$ induced reactions on the $^{40}Ar$ target, for the incoming beam energies of
$E=4$ GeV (left panel) and $E=6$ GeV (right panel).

From Fig.~\ref{fig8}, we observe that the differential cross sections for $\nu_\mu-^{40}Ar$ peak in the intermediate region of $x$, when no cut is applied on the center of 
mass of energy $W$. The differential cross sections decrease with increasing $x$ due to nuclear structure effects. 
The results obtained using the spectral function only (dashed line) are suppressed relative to those for the 
free nucleon case. Quantitatively, this suppression is about $10\%$ at $x=0.3$, $12\%$ at $x=0.6$, and about $6\%$ 
at $x=0.8$ for $E=4$ GeV, when no cut is applied on $W$; for $E=6$ GeV, it ranges from 5-13\% for $0.3\le x\le 0.8$. 
Furthermore, an enhancement is observed when mesonic contributions are included in the  
low-to-intermediate $x~(\lesssim 0.5)$ region; for example, it is about $10-12\%$ at $x=0.3$, $4-5\%$ at $x=0.4$ and
about $1\%$ at $x=0.5$ for $4\le E\le 6$ GeV, in the absence of any $W$ cut. We find that, due to the inclusion of 
nuclear medium effects, there is an enhancement in the differential cross section obtained using the full model in the region $x\le 0.3$
relative to the $\nu_\mu-N$ DIS cross sections for 4 GeV$\le E\le 6$ GeV; for example,
this enhancement is about 16-18\% at $x=0.15$ and $4-5\%$ at $x=0.25$. While for $x>0.3$, there is a reduction of about 8-9\% at $x=0.4$, 12-14\% for $0.5\le x \le 0.6$
and 4-6\% at $x=0.8$ in the energy region 4 GeV$\le E\le 6$ GeV.
% Moreover, the mesonic contribution becomes almost negligible when a $W$ cut is applied. 
% In Fig.~\ref{fig9}, the results of $\frac{1}{E}\;\frac{d\sigma }{dx}$ as a function of $x$ are presented for the $\bar\nu_\mu-^{40}Ar$ DIS process at 
% $E=4$ GeV  and $E=6$ GeV . 

We find that the numerical results for $\bar\nu_\mu-^{40}Ar$ DIS process (Fig.~\ref{fig9}) follow a similar $x$ dependence to that  
observed in the $\nu_\mu-^{40}Ar$ DIS process (Fig.~\ref{fig8}). 
% When mesonic contributions are included, the differential cross sections are significantly enhanced in the  
% low-to-intermediate $x~(\le0.6)$ region. For example, the enhancement is about $66\%$ at $x=0.3$, $42\%$ at $x=0.4$ and 
% about $18\%$ at $x=0.5$ for $E=4$ GeV. This enhancement decreases 
% with increasing beam energy, becoming about 50\%, 31\% and 14\% at $E=6$ GeV for the corresponding values of $x$. We find that the due to the inclusion of 
% nuclear medium effects there is reduction in the differential cross section obtained using the full model which is
% about 11-18\% in the region of $x\le 0.3$, while for $x>0.3$ there is an enhancement of about 10-13\% 
% relative to the case of $\nu_\mu-N$ DIS cross sections for 4 GeV$\le E\le6$ GeV. 
In the case of $\bar\nu_\mu-N$ DIS cross sections, we 
observe that the impact of nuclear medium effects is to enhance the differential cross sections obtained using the full model
for $x\le 0.55$ relative to the free nucleon case. For example, at $x=0.2$, this enhancement is about 75\% (50\%), which becomes
48\% (34\%) at $x=0.3$ and to $4\% (1\%)$ at $x=0.5$ for $E=4(6)$ GeV. However, for $x>0.55$, these effects lead to a reduction in the differential
cross sections, amounting to about 9-14\% in the energy region 4 GeV$\le E \le 6$ GeV. 

Moreover, upon comparing the results in Figs.~\ref{fig8} and \ref{fig9}, it is observed that the differential cross sections for the $\bar\nu_\mu-^{40}Ar$ DIS 
process are significantly smaller than those for the $\nu_\mu-^{40}Ar$ DIS case. Quantitatively, the $\frac{1}{E}\;\frac{d\sigma }{dx}$ results 
obtained using the spectral function only, are suppressed by about 80\%-82\% at $E=4$ GeV and by 74\%-77\% at $E=6$ GeV for
$0.2\le x\le 0.5$, relative to the corresponding neutrino-argon scattering cross sections.
% This implies a smaller number of expected antineutrino events compared to neutrino events. Furthermore, 
% nuclear medium effects are found to be more pronounced in antineutrino-induced processes than in neutrino-induced ones, although they decrease with increasing 
% beam energy. 

From Fig.~\ref{fig8}, it may also be noticed that when a cut of $W \ge$ 2 GeV is applied, the effect of mesonic contribution becomes negligible. 
Furthermore, the peak of the differential cross sections shifts toward lower values of $x$ upon applying the $W$ cut, and the cross section
is further suppressed in the entire $x$ region relative to the case without a cut on $W$.  
From Fig.~\ref{fig9}, it is important to note that the mesonic cloud contribution remains significant even after 
applying a cut of $W\ge 2$ GeV, unlike in the neutrino induced case. Moreover, 
the suppression in the neutrino(antineutrino) induced deep inelastic differential scattering cross sections obtained with the full model (``Total'') due to the application of the $W\ge 2$ GeV cut is about
27\%(70\%) at $x=0.25$ and 63\%(94\%) at $x=0.4$ for $E=4$ GeV, which reduces to 21\%(56\%) and $\sim 40\%$(80\%), respectively, at $E=6$ GeV. Hence, the number of expected antineutrino events in the kinematic region corresponding to safe DIS, i.e., $W\ge 2$ GeV and $Q^2\ge 1$ GeV$^2$, is
significantly suppressed compared to the case without a $W$ cut. This behavior indicates that, in the high-$x$ region corresponding to lower
values of the center of mass energy, the contribution from the nucleon resonance region becomes important. Consequently, the 
applicability of the deep inelastic scattering formalism in the transition region ($W \le$ 2 GeV) may be inadequate, as 
it could lead to double counting of neutrino events and introduce additional uncertainties in the determination of scattering cross sections.

\section{Summary and Conclusion}\label{summary}
In this work, we have studied $\nu_\mu$ and $\bar\nu_\mu$ interactions on $^{40}Ar$ target in the energy region of a few GeV. The numerical results have been 
presented for the differential scattering cross sections off free nucleon as well as off the argon nuclear target in the kinematic region of 
$Q^2\ge 1$ GeV$^2$ without and with different cuts on the center of mass energy $W$.
We summarize our findings below:
\begin{itemize}
\item The nuclear medium effects on the differential cross sections for $\nu_\mu$-$^{40}$Ar and $\bar{\nu}_\mu$-$^{40}$Ar DIS processes, arising 
from the inclusion of spectral functions that incorporate Fermi motion, binding energy, and nucleon correlations, lead to a clear 
suppression of the cross sections. Importantly, this reduction is strongly kinematics dependent and is not uniform across the entire 
range of Bjorken $x$ and $y$, highlighting the nontrivial role of nuclear dynamics in shaping the observed behavior.

\item The mesonic contributions are found to play a decisive role in enhancing the cross sections, particularly in the intermediate region 
of $x$, relative to the results obtained with only nuclear structure effects. This enhancement can be as large as 
$14-20\%$ at $x =0.35$ and $y =0.5$, and remains at a comparable level of
about $7-10\%$ at $x =0.45$ and $y =0.5$ for $\nu_\mu$-induced scattering on the $^{40}$Ar target in the energy region 4 GeV$\le E\le 6$ GeV. In 
the case of antineutrino induced processes, a similarly significant enhancement of $38-48$\% and $16-25$\%
is observed over the same kinematic region of $x$ and $y$, underscoring the crucial impact of mesonic degrees of freedom.

\item The effect of shadowing and antishadowing corrections are very small in the considered kinematic range of $x$ and $y$.

\item The effects of shadowing and antishadowing corrections are found to be negligibly small in the considered kinematic 
range of $x$ and $y$, indicating that their overall impact on the differential cross sections is minimal.

\item The nuclear medium effects are found to be significantly more pronounced in antineutrino-induced processes 
than in neutrino-induced processes off the argon nucleus.

\item The imposition of a cut on the center of mass energy is found to significantly 
suppress the cross section in the considered kinematic region of $x$ and $y$ for $4 \leq E \leq 6$ GeV. Notably, this
suppression becomes increasingly pronounced with rising $x$, while it systematically weakens with increasing beam energy.

\item The reduction in the $\bar{\nu}_\mu$-$^{40}$Ar differential cross sections due to the imposition of a cut on the 
center of mass energy is found to be significantly more substantial than that observed for the $\nu_\mu$-$^{40}$Ar 
differential cross sections, underscoring a stronger sensitivity of antineutrino-induced processes to this kinematic constraint.

\end{itemize}
To conclude, nuclear medium effects are found to play a crucial and non-negligible role in the
kinematic regime relevant to DUNE and SBND energies. These effects have a direct and significant
impact on (anti)neutrino energy reconstruction in ongoing and future experiments, thereby influencing 
the precision of oscillation measurements. Furthermore, a deeper and more systematic investigation of 
nuclear medium effects is essential, particularly in the $Q^2$ region beyond the domain where perturbative 
QCD is reliably applicable. Such studies are indispensable for achieving a comprehensive and accurate 
understanding of deep inelastic lepton-nucleus scattering.

\section{Acknowledgment}   
F. Zaidi is thankful to Council of Scientific \&
Industrial Research, Govt. of India for providing Senior Research Associateship (SRA) under the Scientist’s Pool
Scheme, file no. 13(9240-A)2023-POOL and to the Department of Physics, Aligarh Muslim University, Aligarh for
providing the necessary facilities to pursue this research work. M. S. A. is thankful to the Department of Science
and Technology (DST), Government of India for providing financial assistance under Grant No. SR/MF/PS-01/2016-
AMU/G.


\begin{thebibliography}{100}
\bibitem{DUNE:2018tke}
B.~Abi \textit{et al.} [DUNE],
[arXiv:1807.10334 [physics.ins-det]].

\bibitem{Cicero:2025mab}
V.~Cicero [DUNE],
Nucl. Instrum. Meth. A \textbf{1080}, 170801 (2025).

\bibitem{Hyper-Kamiokande:2025fci}
K.~Abe \textit{et al.} [Hyper-Kamiokande],
Eur. Phys. J. C \textbf{86}, no.2, 170 (2026)
doi:10.1140/epjc/s10052-025-14938-9
[arXiv:2505.15019 [hep-ex]].

\bibitem{ICARUS:2023gpo}
P.~Abratenko \textit{et al.} [ICARUS],
Eur. Phys. J. C \textbf{83}, no.6, 467 (2023)
doi:10.1140/epjc/s10052-023-11610-y
[arXiv:2301.08634 [hep-ex]].

\bibitem{Wood:2024jos}
A.~P.~Wood,
[arXiv:2504.08950 [hep-ex]].


\bibitem{SBND:2025lha}
R.~Acciarri \textit{et al.} [SBND],
[arXiv:2504.00245 [hep-ex]].

\bibitem{Freire:2025hvi}
H.~M.~Freire [SBND],
FERMILAB-MASTERS-2025-02.

\bibitem{MicroBooNE:2025aiw}
P.~Abratenko \textit{et al.} [MicroBooNE],
doi:10.1103/lwrr-98vx
[arXiv:2511.17342 [hep-ex]].

\bibitem{SajjadAthar:2022pjt}
M.~Sajjad Athar, A.~Fatima and S.~K.~Singh,
Prog. Part. Nucl. Phys. \textbf{129}, 104019 (2023).

\bibitem{Choudhary:2026cjh}
B.~Choudhary \textit{et al.} [NOvA],
Springer Proc. Phys. \textbf{432}, 46-51 (2026)
doi:10.1007/978-981-95-1513-4{\_}8

\bibitem{Kalitkina:2025hbg}
A.~Kalitkina [NOvA],
Phys. Atom. Nucl. \textbf{88}, no.3, 443-447 (2025)
doi:10.1134/S1063778825600654

\bibitem{MINERvA:2025hzq}
A.~Lozano \textit{et al.} [MINERvA],
[arXiv:2503.20043 [hep-ex]].

\bibitem{MINERvA:2023ner}
S.~Henry \textit{et al.} [MINERvA],
Phys. Rev. D \textbf{109}, no.9, 092008 (2024)
doi:10.1103/PhysRevD.109.092008
[arXiv:2312.16631 [hep-ex]].

\bibitem{Duffy:2021hie}
K.~E.~Duffy \textit{et al.} [MicroBooNE and ArgoNeuT],
Eur. Phys. J. ST \textbf{230}, no.24, 4275-4291 (2021)
doi:10.1140/epjs/s11734-021-00297-5

\bibitem{ArgoNeuT:2020kir}
R.~Acciarri \textit{et al.} [ArgoNeuT],
Phys. Rev. D \textbf{102}, no.1, 011101 (2020)
doi:10.1103/PhysRevD.102.011101
[arXiv:2004.01956 [hep-ex]].

\bibitem{SajjadAthar:2020nvy}
M.~Sajjad Athar and J.~G.~Morf{\'\i}n,
J. Phys. G \textbf{48}, no.3, 034001 (2021).

\bibitem{GENIE:2021npt}
L.~Alvarez-Ruso \textit{et al.} [GENIE],
Eur. Phys. J. ST \textbf{230}, no.24, 4449-4467 (2021)
doi:10.1140/epjs/s11734-021-00295-7
[arXiv:2106.09381 [hep-ph]].

\bibitem{Bodek:2010km} 
  A.~Bodek and U.~K.~Yang,
    arXiv:1011.6592 [hep-ph];  Nucl.\ Phys.\ Proc.\ Suppl.\  {\bf 112}, 70 (2002).
    
    

\bibitem{EuropeanMuon:1983wih}
J.~J.~Aubert \textit{et al.} [European Muon],
Phys. Lett. B \textbf{123}, 275-278 (1983).

\bibitem{Arneodo:1992wf}
M.~Arneodo,
Phys. Rept. \textbf{240}, 301-393 (1994).

\bibitem{Gomez:1993ri}
J.~Gomez, R.~G.~Arnold, P.~E.~Bosted, C.~C.~Chang, A.~T.~Katramatou, G.~G.~Petratos, A.~A.~Rahbar, S.~E.~Rock, A.~F.~Sill and Z.~M.~Szalata, \textit{et al.}
Phys. Rev. D \textbf{49}, 4348-4372 (1994).

\bibitem{SajjadAthar:2025fhk}
M.~Sajjad Athar,
Springer Proc. Phys. \textbf{432}, 20-26 (2026).

\bibitem{Hirai:2007sx}
M.~Hirai, S.~Kumano and T.~H.~Nagai,
Phys. Rev. C \textbf{76}, 065207 (2007)
doi:10.1103/PhysRevC.76.065207
[arXiv:0709.3038 [hep-ph]].

\bibitem{Eskola:2009uj}
K.~J.~Eskola, H.~Paukkunen and C.~A.~Salgado,
JHEP \textbf{04}, 065 (2009)
doi:10.1088/1126-6708/2009/04/065
[arXiv:0902.4154 [hep-ph]].

\bibitem{deFlorian:2011fp}
D.~de Florian, R.~Sassot, P.~Zurita and M.~Stratmann,
Phys. Rev. D \textbf{85}, 074028 (2012)
doi:10.1103/PhysRevD.85.074028
[arXiv:1112.6324 [hep-ph]].

\bibitem{Kovarik:2012zz}
K.~Kovarik, I.~Schienbein, T.~Stavreva, F.~I.~Olness, J.~Y.~Yu, C.~Keppel, J.~G.~Morfin and J.~F.~Owens,
Few Body Syst. \textbf{52}, 271-277 (2012)
doi:10.1007/s00601-011-0297-7

\bibitem{Kovarik:2015cma}
K.~Kovarik, A.~Kusina, T.~Jezo, D.~B.~Clark, C.~Keppel, F.~Lyonnet, J.~G.~Morfin, F.~I.~Olness, J.~F.~Owens and I.~Schienbein, \textit{et al.}
Phys. Rev. D \textbf{93}, no.8, 085037 (2016)
doi:10.1103/PhysRevD.93.085037
[arXiv:1509.00792 [hep-ph]].

\bibitem{private} J. G. Morfin, private communication.








































































\bibitem{Marco:1995vb} 
  E.~Marco {\it et al.}, 
    Nucl.\ Phys.\ A {\bf 611}, 484 (1996).
  
  

\bibitem{FernandezdeCordoba:1991wf}
  P.~Fernandez de Cordoba and E.~Oset,
    Phys. Rev. C {\bf 46}, 1697 (1992).
    
    

\bibitem{Kulagin:2004ie} 
  S.~A.~Kulagin and R.~Petti,
    Nucl.\ Phys.\ A {\bf 765}, 126 (2006).
  
  

\bibitem{Harland-Lang:2014zoa} 
  L.~A.~Harland-Lang, A.~D.~Martin, P. Motylinski and R.~S.~Thorne,
    Eur.\ Phys.\ J.\ C {\bf 75}, no. 5, 204 (2015).   
    

\bibitem{Vermaseren:2005qc} 
  J.~A.~M.~Vermaseren {\it et al.}, 
    Nucl.\ Phys.\ B {\bf 724}, 3 (2005).
    
              

\bibitem{Moch:2004xu} 
S.~Moch, J.~A.~M.~Vermaseren and A.~Vogt,
Phys.\ Lett.\ B {\bf 606}, 123 (2005).

\bibitem{Moch:2008fj} 
  S.~Moch, J.~A.~M.~Vermaseren and A.~Vogt,
    Nucl.\ Phys.\ B {\bf 813}, 220 (2009).
   
           

\bibitem{Schienbein:2007gr}
  I.~Schienbein {\it et al.},
    J.\ Phys.\ G {\bf 35}, 053101 (2008).

\bibitem{Zaidi:2019asc}
F.~Zaidi, H.~Haider, M.~Sajjad Athar, S.~K.~Singh and I.~Ruiz Simo,
Phys. Rev. D \textbf{101}, no.3, 033001 (2020).

\bibitem{halzen_martin} F. Halzen and A.D. Martin
  ``Quarks and Leptons: An Introductory Course in Modern Particle Physics ",
  Pg. no. 207, Ed. John Wiley \& Sons, Reprint (2008).  

\bibitem{Callan:1969uq} 
  C.~G.~Callan, Jr. and D.~J.~Gross,
    Phys.\ Rev.\ Lett.\  {\bf 22}, 156 (1969).
  
    
              

\bibitem{NNPDF:2014otw}
R.~D.~Ball \textit{et al.} [NNPDF],
JHEP \textbf{04}, 040 (2015).

\bibitem{Bailey:2020ooq}
S.~Bailey, T.~Cridge, L.~A.~Harland-Lang, A.~D.~Martin and R.~S.~Thorne,
Eur. Phys. J. C \textbf{81}, no.4, 341 (2021).

\bibitem{Yan:2022pzl}
M.~Yan, T.~J.~Hou, P.~Nadolsky and C.~P.~Yuan,
Phys. Rev. D \textbf{107}, no.11, 116001 (2023).


\bibitem{Gluck:2007ck}
M.~Gluck, P.~Jimenez-Delgado and E.~Reya,
Eur. Phys. J. C \textbf{53}, 355-366 (2008).


\bibitem{Lai:2009ne}
H.~L.~Lai, J.~Huston, S.~Mrenna, P.~Nadolsky, D.~Stump, W.~K.~Tung and C.~P.~Yuan,
JHEP \textbf{04}, 035 (2010).

\bibitem{Alekhin:2018pai}
S.~Alekhin, J.~Bl{\"u}mlein and S.~Moch,
Eur. Phys. J. C \textbf{78}, no.6, 477 (2018).

\bibitem{Zaidi:2019mfd} 
  F.~Zaidi, H.~Haider, M.~Sajjad Athar, S.~K.~Singh and I.~Ruiz Simo,
    Phys.\ Rev.\ D {\bf 99}, no. 9, 093011 (2019).
  
    

\bibitem{Baym:1975vb} 
  G.~Baym and G.~E.~Brown,
    Nucl.\ Phys.\ A {\bf 247}, 395 (1975).
  

\bibitem{Gluck:1991ey}
  M.~Gluck {\it et al.}, 
    Z.\ Phys.\  C {\bf 53}, 651 (1992).
  
  









































































% 

% %\cite{Ansari:2021cao}

% \bibitem{Ansari:2021cao}

% V.~Ansari, M.~S.~Athar, H.~Haider, I.~Ruiz Simo, S.~K.~Singh and F.~Zaidi,

% %``Deep inelastic (anti)neutrino{\textendash}nucleus scattering,''

% Eur. Phys. J. ST \textbf{230}, no.24, 4433-4448 (2021).

% % doi:10.1140/epjs/s11734-021-00277-9

% % [arXiv:2106.14670 [hep-ph]].

% %11 citations counted in INSPIRE as of 11 Apr 2026

% 

%``High-energy electroproduction and the constitution of the electric current,''

%\cite{EuropeanMuon:1983wih}

%``The ratio of the nucleon structure functions $F2_n$ for iron and deuterium,''

%\cite{Gomez:1993ri}

%``Measurement of the A-dependence of deep inelastic electron scattering,''

% doi:10.1103/PhysRevD.49.4348

%480 citations counted in INSPIRE as of 18 Mar 2026

%\cite{Arneodo:1992wf}

%``Nuclear effects in structure functions,''

% doi:10.1016/0370-1573(94)90048-5

%624 citations counted in INSPIRE as of 18 Mar 2026

%\cite{Lai:2009ne}

%``Parton Distributions for Event Generators,''

% doi:10.1007/JHEP04(2010)035

% [arXiv:0910.4183 [hep-ph]].

%71 citations counted in INSPIRE as of 18 Mar 2026

%\cite{Alekhin:2018pai}

%``NLO PDFs from the ABMP16 fit,''

% doi:10.1140/epjc/s10052-018-5947-1

% [arXiv:1803.07537 [hep-ph]].

%109 citations counted in INSPIRE as of 18 Mar 2026

%\cite{Gluck:2007ck}

%``Dynamical parton distributions of the nucleon and very small-x physics,''

% doi:10.1140/epjc/s10052-007-0462-9

% [arXiv:0709.0614 [hep-ph]].

%255 citations counted in INSPIRE as of 18 Mar 2026

% doi:10.1016/0370-2693(83)90437-9

%1686 citations counted in INSPIRE as of 18 Mar 2026

%\cite{SajjadAthar:2022pjt}

%``Neutrinos and their interactions with matter,''

% doi:10.1016/j.ppnp.2022.104019

% [arXiv:2206.13792 [hep-ph]].

%61 citations counted in INSPIRE as of 05 Mar 2026

%\cite{SajjadAthar:2025fhk}

%``Neutrino Interactions in~the~SIS and~DIS Regions: Current Insights and~Future Challenges,''

% doi:10.1007/978-981-95-1513-4{\_}4

% [arXiv:2502.16107 [hep-ph]].

%0 citations counted in INSPIRE as of 26 Feb 2026

%\cite{GENIE:2021npt}

%``Recent highlights from GENIE v3,''

%132 citations counted in INSPIRE as of 06 Mar 2026

%\cite{MicroBooNE:2025aiw}

%``Measurements of differential charged-current cross sections on argon for electron neutrinos with final-state protons in MicroBooNE,''

%0 citations counted in INSPIRE as of 06 Mar 2026

%\cite{Zaidi:2019asc}

%``Weak structure functions in $\nu_l-N$ and $\nu_l-A$ scattering with nonperturbative and higher order perturbative QCD effects,''

% doi:10.1103/PhysRevD.101.033001

% [arXiv:1911.12573 [hep-ph]].

% %21 citations counted in INSPIRE as of 22 Nov 2025

%\cite{SajjadAthar:2020nvy}

%``Neutrino(antineutrino){\textendash}nucleus interactions in the shallow- and deep-inelastic scattering regions,''

% doi:10.1088/1361-6471/abbb11

% [arXiv:2006.08603 [hep-ph]].

%62 citations counted in INSPIRE as of 05 Mar 2026

%\cite{Kalitkina:2025hbg}

%``NOvA Recent Results of Three-Flavor Oscillation Analysis,''

%1 citations counted in INSPIRE as of 06 Mar 2026

%\cite{MINERvA:2025hzq}

%``Measurement of charged-current $\nu_\mu$ and $\bar{\nu}_\mu$ cross sections on hydrocarbon in a shallow inelastic scattering region,''

%0 citations counted in INSPIRE as of 06 Mar 2026

%\cite{MINERvA:2023ner}

%``Measurement of electron neutrino and antineutrino cross sections at low momentum transfer,''

%7 citations counted in INSPIRE as of 06 Mar 2026

%\cite{Duffy:2021hie}

%``Neutrino interaction measurements with the MicroBooNE and ArgoNeuT liquid argon time projection chambers,''

%4 citations counted in INSPIRE as of 06 Mar 2026

%\cite{ArgoNeuT:2020kir}

%``First measurement of electron neutrino scattering cross section on argon,''

%32 citations counted in INSPIRE as of 06 Mar 2026

%\cite{Hirai:2007sx}

%``Determination of nuclear parton distribution functions and their uncertainties in next-to-leading order,''

%446 citations counted in INSPIRE as of 05 Mar 2026

%\cite{Eskola:2009uj}

%``EPS09: A New Generation of NLO and LO Nuclear Parton Distribution Functions,''

%1344 citations counted in INSPIRE as of 05 Mar 2026

%\cite{deFlorian:2011fp}

%``Global Analysis of Nuclear Parton Distributions,''

%339 citations counted in INSPIRE as of 05 Mar 2026

%\cite{Kovarik:2012zz}

%``Nuclear corrections in nu A DIS and their compatibility with global NPDF analyses,''

%5 citations counted in INSPIRE as of 05 Mar 2026

%\cite{Kovarik:2015cma}

%``nCTEQ15 - Global analysis of nuclear parton distributions with uncertainties in the CTEQ framework,''

%666 citations counted in INSPIRE as of 05 Mar 2026

%\cite{AtharSajjad:2022ipr}

%``Nuclear medium effects in lepton-nucleus DIS in the region of x{\ensuremath{\gtrsim}}1,''

% doi:10.1103/PhysRevD.105.093002

% [arXiv:2202.11892 [hep-ph]].

%4 citations counted in INSPIRE as of 02 Mar 2026

%\cite{SBND:2025lha}

%``The Short-Baseline Near Detector at Fermilab:~Input to the European Strategy for Particle Physics 2026 Update,''

%9 citations counted in INSPIRE as of 06 Mar 2026

%\cite{Freire:2025hvi}

%``Selection Algorithm for electron neutrino charged current interactions in SBND,''

%0 citations counted in INSPIRE as of 06 Mar 2026

%\cite{Hyper-Kamiokande:2025fci}

%``Sensitivity of the Hyper-Kamiokande experiment to neutrino oscillation parameters using accelerator neutrinos,''

%18 citations counted in INSPIRE as of 06 Mar 2026

%\cite{Choudhary:2026cjh}

%``Latest Results from~the~NOvA Experiment,''

%0 citations counted in INSPIRE as of 06 Mar 2026

%\cite{DUNE:2018tke}

%``The DUNE Far Detector Interim Design Report Volume 1: Physics, Technology and Strategies,''

%277 citations counted in INSPIRE as of 26 Feb 2026

%\cite{Cicero:2025mab}

%``Imaging neutrino interactions with liquid Argon scintillation light at the DUNE near detector complex,''

% doi:10.1016/j.nima.2025.170801

%2 citations counted in INSPIRE as of 26 Feb 2026

%\cite{ICARUS:2023gpo}

%``ICARUS at the Fermilab Short-Baseline Neutrino program: initial operation,''

%74 citations counted in INSPIRE as of 06 Mar 2026

%\cite{Wood:2024jos}

%``The NuMI Neutrino Flux Prediction at ICARUS,''

%0 citations counted in INSPIRE as of 06 Mar 2026

%\cite{Zaidi:2021iam}

%``Nuclear medium effects in the deep inelastic $\nu_\tau/\bar\nu_\tau-^{40}Ar$ scattering at DUNE energies,''

% doi:10.1103/PhysRevD.105.033010

% [arXiv:2111.07609 [nucl-th]].

% %8 citations counted in INSPIRE as of 02 Mar 2026

%\cite{NNPDF:2014otw}

%``Parton distributions for the LHC Run II,''

% doi:10.1007/JHEP04(2015)040

% [arXiv:1410.8849 [hep-ph]].

%4405 citations counted in INSPIRE as of 21 Jan 2026

%\cite{Bailey:2020ooq}

%``Parton distributions from LHC, HERA, Tevatron and fixed target data: MSHT20 PDFs,''

% doi:10.1140/epjc/s10052-021-09057-0

% [arXiv:2012.04684 [hep-ph]].

%588 citations counted in INSPIRE as of 21 Jan 2026

%\cite{Yan:2022pzl}

%``CT18 global PDF fit at leading order in QCD,''

% doi:10.1103/PhysRevD.107.116001

% [arXiv:2205.00137 [hep-ph]].

%17 citations counted in INSPIRE as of 21 Jan 2026

\end{thebibliography}
\end{document}